# Mind the Gap – Imaging Buried Interfaces in Twisted Oxide Moirés

*Harikrishnan KP[1,*], Xin Wei[2,3], Chia-Hao Lee[1], Dasol Yoon[4], Yonghun Lee[2,5], Kevin J. Crust[2,3], Yu-Tsun Shao[1,6], Ruijuan Xu[2,5,7], Jong-Hoon Kang[8,9,10], Ce Liang[11], Jiwoong Park[8,11,12], Harold Y. Hwang[2,5], David A. Muller[1,13,*]*

[1] *School of Applied and Engineering Physics, Cornell University, Ithaca, NY 14853, USA.*

[2] *Stanford Institute for Materials and Energy Sciences, SLAC National Accelerator Laboratory, Menlo Park, CA 94025, USA.*

[3] *Department of Physics, Stanford University, Stanford, CA 94305, USA.*

[4] *Department of Materials Science and Engineering, Cornell University, Ithaca, NY 14853, USA.*

[5] *Department of Applied Physics, Stanford University, Stanford, CA 94305, USA.*

[6] *Mork Family Department of Chemical Engineering and Materials Science, University of Southern California, Los Angeles, CA 90089, USA.*

[7] *Department of Materials Science and Engineering, North Carolina State University, Raleigh, NC 27695, USA.*

[8] *Department of Chemistry, University of Chicago, Chicago, IL 60637, USA.*

[9] *Center for Van der Waals Quantum Solids, Institute for Basic Science (IBS), Pohang 37673, Korea.*

[10] *Department of Materials Science and Engineering, Pohang University of Science and Technology (POSTECH), Pohang 37673, Korea.*

[11] *Pritzker School of Molecular Engineering, University of Chicago, IL 60637, USA.*

[12] *James Franck Institute, University of Chicago, Chicago, IL 60637, USA.*

[13] *Kavli Institute at Cornell for Nanoscale Science, Cornell University, Ithaca, NY 14853, USA.*

[*] *Corresponding Authors: E-mail: hk944@cornell.edu, david.a.muller@cornell.edu*

**Abstract**

The ability to tune electronic structure in twisted stacks of two-dimensional (2D) materials has motivated the exploration of similar moiré physics with twisted oxide membranes. Due to the intrinsic three-dimensional nature of bonding in many oxides, achieving atomic-level coupling is significantly more challenging than with van der Waals materials. Although clean interfaces with atomic-level proximity have been demonstrated in ceramic bicrystals using high-temperature and high-pressure processing to facilitate atomic diffusion that flattens rough interfaces, such conditions are not readily accessible when bonding oxide membranes. This study shows how topographic mismatch due to surface roughness of the membranes can restrict atomic-scale proximity at the interface to isolated patches even after contaminants and amorphous interlayers are eliminated. In interfaces between 2D materials and oxide membranes the reduced ability of the 2D material to conform to the membrane's step-terrace topography also limits atomic-scale contact. When imaging stacked membranes in projection, we find conventional through-focal imaging to be relatively insensitive to the buried interface, whereas electron ptychography detects structural variations on the order of a nanometer. These findings highlight interface roughness as a key challenge for the field of oxide twistronics and emphasize the need for reliable characterization methods, both in cross-section and projection.

Keywords: oxide membranes, moiré heterostructures, electron ptychography, interfacial coupling, 3D imaging, 2D-3D heterostructures

## 1. Introduction

Propelled by the experimental discovery of superconductivity in twisted bilayer graphene[1], the last few years have witnessed tremendous interest in the field of "twistronics" where the twist angle between 2D materials in a van der Waals heterostructure is tuned to control electronic correlations in the emergent moiré superlattice[2–5]. In addition to superconductivity, the twist-angle tuning parameter has also led to the demonstration of other correlated electronic states like Mott insulators, Wigner crystals, Chern insulators, and orbital magnetism in these moiré materials constructed with 2D materials as building blocks[6–11]. Recent developments in the growth and manipulation of free-standing oxide membranes have opened up the opportunity to create and explore similar moiré physics in twisted stacks of oxides[12–18]. However, due to the three-dimensional nature of bonding in many oxide materials, achieving atomic scale interlayer electronic coupling is more challenging than in 2D materials. The presence of dangling bonds at the surfaces of oxides can result in surface reconstructions and adsorption, often resulting in a substantial dead layer between the two membranes in a stacked bilayer. Due to the possibility of such an interfacial dead layer from fabrication, investigation of interlayer electronic coupling in stacked bilayer oxides requires careful examination of the close structural proximity of the two layers at the interface. To set a formal meaning to structural proximity thresholds, we use Alvarez's survey of inter-atomic distances[19]. From this survey of over 600,000 compounds which includes over five million "non-bonded" interatomic distances and over a million bond distances, chemical bond distances were determined to be in the range of 2-3 Å while van-der Waals distances were determined to be in the range of 3-5 Å generally, and 4-5 Å for oxygen-metal van der Waals bonds.

Even when interfacial dead layers are eliminated through optimization of the assembly process, here we observe that surface roughness of the membranes hinders interfacial coupling over extended areas, limiting atomic-scale proximity to isolated patches. Interfacial electronic coupling over extended areas have been achieved in fused ceramic bicrystals[20–22] with atomically-sharp interfaces between different oxide single crystals or thin films. Here high temperatures, characteristic of step-flow growth are required to facilitate atomic flow and bridge the interfacial gap. Reproducing the same level of control with membranes is a central challenge for oxide twistronics, and a method to assess interface quality would greatly aid progress in the field.

Due to the nanometer length-scale of the moiré lattice and spatial inhomogeneities in a stacked bilayer sample, scanning transmission electron microscopy (STEM) is well-suited for structural characterization of these stacked heterostructures. However, characterization solely in projection can be misleading - stacked membranes may generate moiré patterns even when separated by a large interfacial gap, and the presence of a projected moiré should therefore not be interpreted as evidence of atomic-scale proximity. The typical approach for 3D imaging using conventional STEM techniques involves acquiring an optical depth-sectioning series (through-focal series) of images by varying the focal point of the probe to different positions in the depth direction. However, this approach is dose-inefficient and can produce inaccurate results for crystalline samples due to strong multiple scattering effects for the electron beam[23,24]. For these reasons, here we show that through-focal imaging with conventional STEM methods is inadequate for detecting and probing dead layers, if present at the interface between stacked membranes. Instead, we apply a new 3D imaging technique - multislice electron ptychography (MEP)[25,26], that enables the accurate reconstruction of the 3D specimen potential[27–33] with a typical depth blur of 2-3 nm. We demonstrate that MEP can detect the presence or absence of strain fields in the membranes as well as interfacial dead layers in stacked bilayers and explore its sensitivity and depth resolution. Our simulations indicate that while the quantitative determination with MEP of a gap between the membranes is possible down to the scale of the depth blur, perhaps surprisingly, the existence of a gap can be detected down to much smaller distances - about 0.5 nm for a typical MEP experiment, even when quantification is no longer reliable.

In general, to demonstrate atomic-abruptness, cross-sectional images are also useful, although here too, it should be remembered that a cross-sectioned image is still a projection, and interpreting the image contrast to properly account for possible, projected in-plane roughness takes some care. In comparing MEP with cross-sectional images, we also identify ~10% systematic underestimation of sample extent in the depth direction in MEP reconstructions, likely arising from the treatment of electron-phonon scattering with an absorptive model.

## 2. Retrieving 3D information using the Parallax Effect

For heterostructures formed by stacked membranes, a characterization method to assess interface quality without the need to prepare a cross-sectional foil is highly desirable. Such a technique should ideally indicate whether an interfacial gap is present and, if absent, be capable of detecting (and resolving in depth) interfacial structural features. In the following sections, we will present experimental results demonstrating the capability of MEP for such 3D characterization and its ability to address some of the limitations of conventional through-focal methods. In this section, we will explain the working principles of the MEP technique that make this characterization possible.

MEP uses the same experimental setup as conventional STEM imaging where a convergent electron beam is scanned across the sample, but now with the momentum-resolved scattering information recorded in the far-field using a pixelated detector for each scan position as shown in Fig. 1(a). At the convergence angles typically used for atomic-resolution imaging, the diffraction pattern consists of overlapping disks, with phase information encoded in their overlap region from the interference of scattered wavefunctions. As the probe position is shifted, this phase information evolves according to the Fourier shift theorem, allowing its extraction from heavily oversampled scans that offer sufficient redundancy for a reliable reconstruction. With the knowledge of a forward model that describes the scattering physics, the specimen potential is determined by iterative algorithms that solve the phase problem with optimization techniques. We use the well-established multislice algorithm[34] as our forward model, where the sample potential is divided into a series of slices along the beam direction, and the electron wavefunction sequentially scatters off each slice with free-space propagation between the slices.

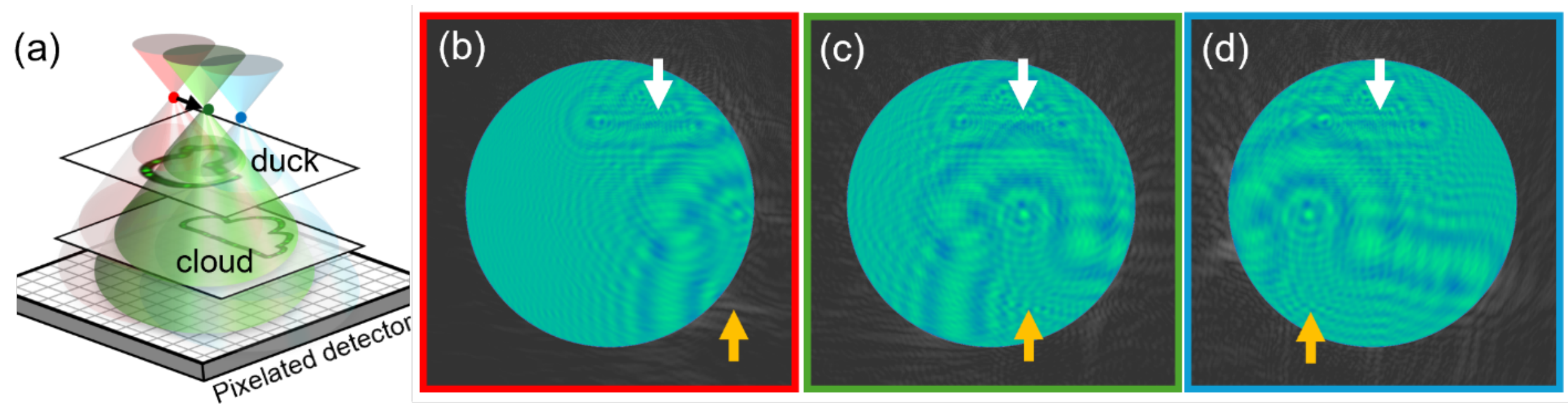

***Figure 1*. How depth information is encoded in ptychographic data.** *(a) Schematic of the experimental setup for ptychography where a converged beam of electrons scatters off the sample and a pixelated detector collects the diffraction pattern in the far-field. The sample here is an array of atoms forming the outline of a duck and a cloud, spatially separated along the beam direction. The diffraction patterns corresponding to three different probe positions indicated with red, green and blue dots are shown in (b-d). Due to the parallax effect, the shadow image of the duck which is closer to the focal point of the probe moves faster across the bright field disk compared to the cloud. The white and yellow arrows in (b-d) point to the center of the cloud and the eye of the duck respectively and acts as a guide to the eye for tracking their movement across the bright field disk. Hence, depth information is encoded in diffraction patterns in the form of differences in distance moved by the scattering imprints from different objects for a fixed shift in probe position.*

As MEP solves the inverse multislice imaging problem, the reconstruction contains multiple slices of the sample potential, reflecting the changes in sample structure along the depth direction. The depth-resolved information is extracted from a single dataset, without the need for a through-focal series and instead relies on the parallax effect[27,35] summarized in Fig. 1. For illustration purposes, we use an array of atoms shaped like a duck and a cloud as two scattering species, with the former located closer to the focal point of the probe. The diffraction patterns calculated for three different probe positions indicated in the schematic are shown in Fig. 1(b-d) and show the imprint of the scattering from the two objects as a shadow image in the bright field disk. The shadow image of the duck moves faster across the bright field disk with shifts in probe position in comparison to the cloud. This observation is explained by the parallax effect, which causes shadow images of objects located closer to the focal point of the probe to move faster in the diffraction plane as the probe is scanned. Hence, depth information is encoded in diffraction patterns in the form of differences in distance moved by the scattering imprints from different objects for a fixed shift in probe position. This information is utilized in MEP reconstructions to achieve depth resolution.

## 3. Buried interfaces in twisted oxide stacks

We use MEP for the three-dimensional imaging of oxide membrane heterostructures and compare the results with depth sectioning using conventional high-angle annular dark field (HAADF) imaging. The experimental conditions and algorithm settings of MEP are summarized as Supplementary Table 1. We show that plan-view through-focal imaging can fail to detect an interfacial dead layer due to channeling and projection artifacts. While MEP is sensitive to the dead layer under certain conditions as demonstrated below, cross-sectional imaging is still essential for examining the interface and the potential for interlayer coupling.

We use an example of a stacked bilayer of $SrTiO_3$ (Sample 1), assembled with typical protocols used in literature[12,13,36] (see Methods section in the Supplementary Information for details on sample preparation). A cross-sectional HAADF image of the stacked bilayer of $SrTiO_3$ with a 9-degree twist between the two membranes is shown in Fig. 2(a), showing a dead layer that appears as a “gap” of ~1.5 nm between the two membranes, similar to prior observations[16,36,37]. We measure a mean spacing of $1.5 \pm 0.3$ nm with a maximum and minimum spacings of 1.8 nm and 0.9 nm respectively (see Supplementary Fig. 1). Changes in the spacing correlate with terrace structures on the membranes. We note, however, that the cross-sectional image is itself a projection, and that projected roughness can make the interfacial gap appear thinner than it actually is (see the Discussion section and Fig. 7). Elemental identification with energy-dispersive X-ray (EDX) spectroscopy indicates that the gap region in sample 1 contains organic materials as well as oxides of residual elements from the sacrificial layer used during growth of the membranes (Supplementary Fig. 2). In order to explore if the dead layer can be removed by optimizing assembly and with higher temperature annealing, we prepared another $SrTiO_3$ bilayer sample - Sample 2 (details of preparation in Methods section). Cross-sectional HAADF imaging of Sample 2 reveals isolated patches (few nm in extent) where the two membranes are in close proximity, closer to bridging the interfacial gap, as shown in Fig. 2(b). Such atomically proximal regions occur only in patches due to the surface roughness of the membranes and their inability to conform to the surface step-terrace structure. We measure a mean projected spacing of $1.7 \pm 0.6$ nm with maximum and minimum projected spacings of 2.5 nm and 0.5 nm respectively (see Supplementary Fig. 1). In Supplementary Fig. 3, we similarly show how the step terraces on the membrane surfaces result in a wide range (0.7 nm – 2.2 nm) for the extent of the interfacial gap, even within a small field of view (~85 nm). EDX elemental profiles across the interface of Sample 2 are also shown in Supplementary Fig. 4. A comparison of plan-view HAADF and MEP images of Sample 2 showing the projected moiré pattern is provided in Supplementary Fig. 5.

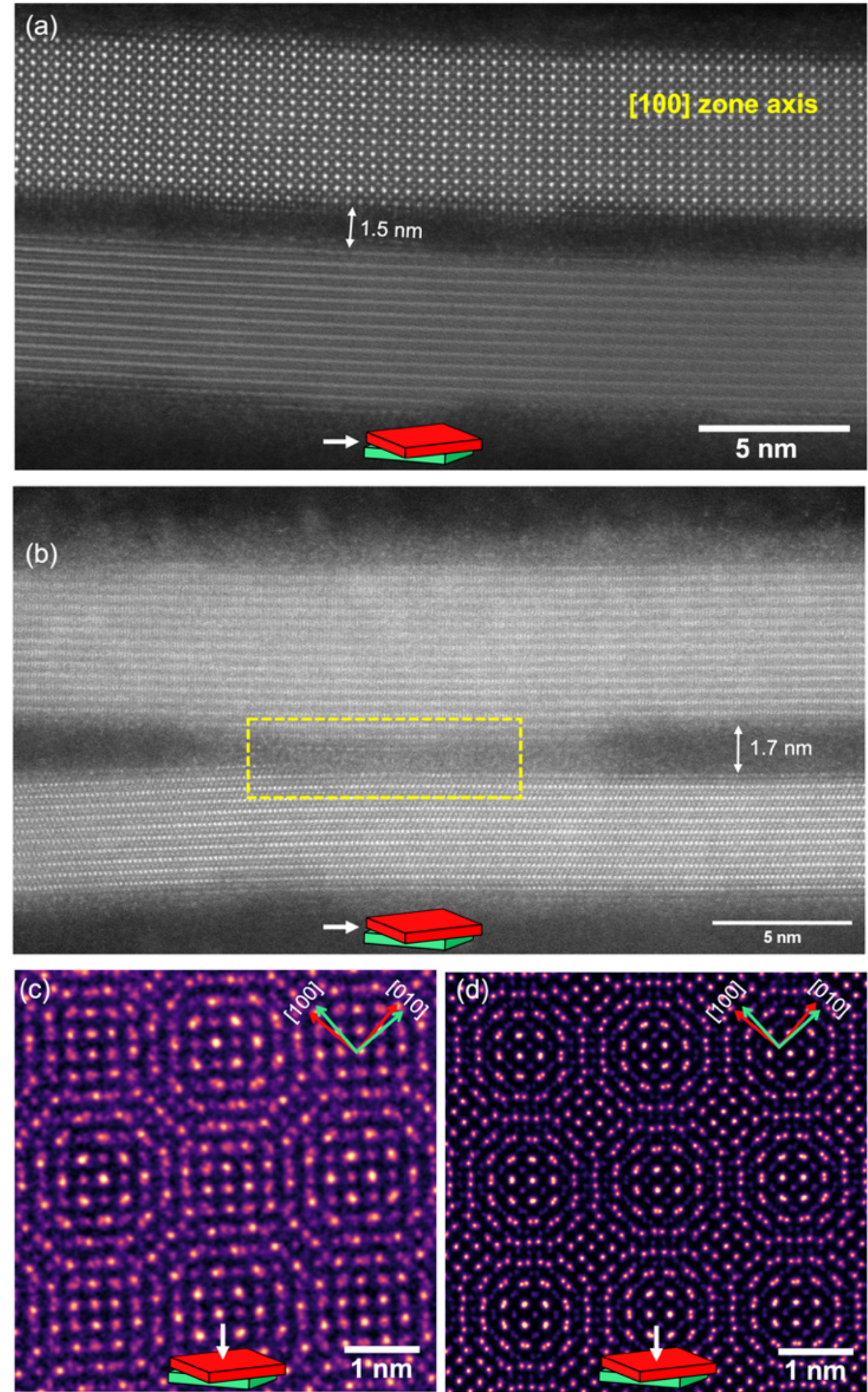

***Figure 2*. *Cross-sectional and plan-view imaging of the stacked bilayer*.** *(a) Cross-sectional HAADF image of stacked bilayer $SrTiO_3$ membranes (Sample 1) showing a dead layer of ~1.5 nm at the interface, suggesting negligible coherent interlayer electronic coupling. (b) Cross-sectional HAADF image of Sample 2 showing an isolated patch labelled with a yellow dotted box, where atomic planes close the interfacial gap. The presence of such patches indicate that atomically proximal interfaces are achievable but eliminating surface roughness is important for getting clean interfaces over extended areas. (c, d) Plan-view projection images of the Sample 1 projected through sample depth, obtained by summing up (c) a through-focal series of HAADF images and (d) all slices in a MEP reconstruction. Both images show the moiré structure arising from the twist angle (9°) between the two membranes. The directions corresponding to the cubic axes for the two membranes are labelled on the top right. Although both images have the same sample orientation and field of view, the HAADF image appears tilted compared to the MEP image as HAADF imaging picks up most contrast from the upper layer, while the MEP image captures contrast uniformly from both layers. The MEP image also shows the oxygen sublattice which is not visible in the HAADF image.*

These findings indicate that while an interface with atomic-level proximity can be achieved over atomic length scales, replication over an extended area requires a clean and smooth surface termination with well-separated terraces for both membranes. While these results show that there are both challenges and opportunities to fabricate atomically precise oxide bilayers, we focus here on Sample 1 and the imaging challenges of a gapped interface.

Even in the presence of a considerable interfacial separation, the stacked bilayer will still produce a moiré pattern when imaged top-down (plan-view) in projection as shown in Fig. 2(c, d). The HAADF-STEM projection image in Fig. 2(c) is obtained by summing over all the images acquired through an optical depth-sectioning series, while Fig. 2(d) shows the projected ptychographic image obtained by summing up all slices from a MEP reconstruction. Although the projection images with the 2 methods shown in Fig. 2(c, d) have the same sample orientation and cover the same spatial extent, they appear visually distinct due to the different dependence of HAADF and multislice ptychographic image intensity on the atomic number (Z) of the scattering atoms. The moiré pattern in the HAADF image also appears tilted relative to that in the MEP image, as the through-focal HAADF series predominantly captures contrast from the upper membrane, whereas the MEP image captures contrast uniformly from both layers.

We additionally note that rescaling the HAADF moiré pattern in both lateral dimensions by a factor of $1/\sqrt{2}$ followed by a 45° rotation makes the resulting pattern visually identical to the moiré pattern obtained with MEP, as shown in Supplementary Fig. 6. The explanation for this connection that holds for the [100] zone axis, as well as schematic models differentiating the AA and AB stacked regions in the moiré pattern, are shown in Supplementary Figs. 7, 8.

Further, we compare the 3D volumetric data obtained using through-focal HAADF series and MEP in Fig. 3, which shows the 3D volume rendering of the twisted bilayer structure when viewed along three orthogonal axes. Prior work has shown that the contrast of HAADF imaging is highly depth-sensitive due to electron beam channeling, with optimal contrast when the probe is focused near the top of the sample, and a gradual decline in contrast as the probe is focused deeper into the sample[38]. Such depth-dependent contrast and limited depth resolution of through-focal HAADF imaging makes it very difficult to detect the gap between the two layers, as shown in Fig. 3(a). Similar channeling artifacts are also observed in integrated Differential Phase Contrast (iDPC) imaging as shown in Supplementary Fig. 9.

In contrast, MEP solves the channeling problem, providing good lateral information transfer throughout the volume of the sample as shown in Fig. 3(b). The enhanced depth resolution of the method reveals the presence of the amorphous interlayer between the two membranes. The atomic columns in each layer are uncoupled and undistorted, indicating the absence of any interlayer structural coupling. Individual slices at different depths showing the variations in the sample structure along the beam direction are shown in Supplementary Fig. 10.

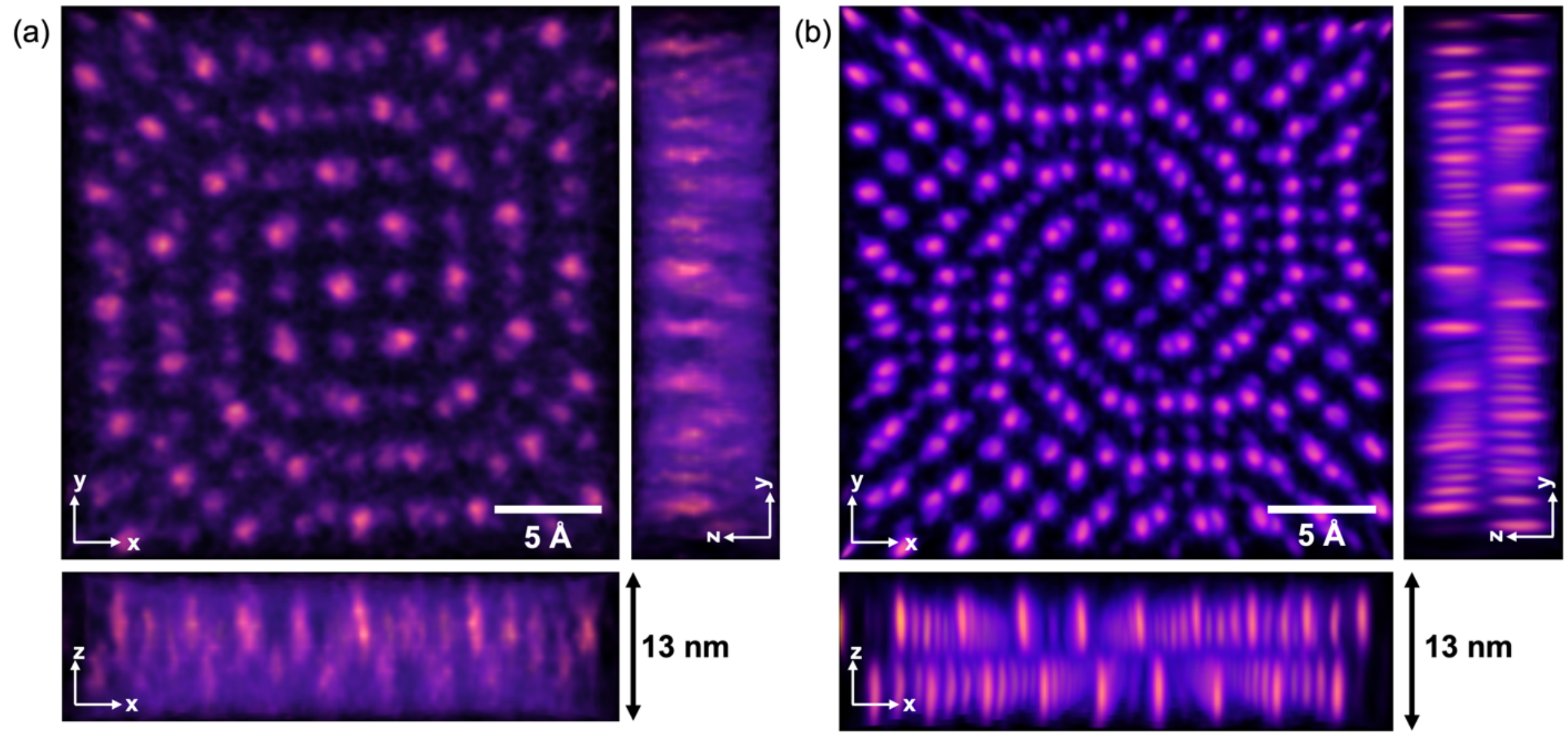

***Figure 3**. **3D visualization of the twisted, stacked bilayer** with (a) HAADF and (b) multislice ptychography. The 3D volume obtained by optical depth sectioning with HAADF is unable to resolve the gap, due to limited depth resolution and its contrast depending heavily on electron beam channeling. The 3D reconstruction with multislice ptychography shows the gap with uncoupled and undistorted atomic columns in the 2 layers, indicating the absence of structural coupling.*

Being a computational imaging method, MEP is sensitive to the choice of reconstruction parameters, including constraints and regularizations that are applied to achieve a stable convergence. For example, enforcing a heavy regularization in the depth direction[26,27,39] during MEP reconstructions can blur the interfacial gap as shown in Supplementary Fig. 11. Moreover, the enhanced mixing between adjacent slices due to such a heavy regularization can create artificial structural distortions near the interface that can propagate farther into the two layers. Therefore, cross-sectional imaging is essential to validate and constrain the interface atomic structure.

## 4. Benchmarking depth resolution of MEP

The depth-sectioning capability of MEP is further explored in Fig. 4, where we plot depth profiles of individual atomic columns. Figure 4(a) shows the projected MEP reconstruction centered on an AA stacked region (local region where Sr, Ti and O columns on the 2 layers are aligned) of the moiré pattern, where three different positions are labelled. The depth profiles of the atomic columns at these labelled positions are plotted in Fig. 4(b) and clearly indicate that the column labelled in red (green) is confined to the top (bottom) membrane. The depth profile for the position marked in cyan contains atomic columns in both layers and shows a considerable dip in intensity at the interface, indicating the presence of the gap between the two membranes. The measured minima-to-peak ratio of 0.59 surpasses that needed to meet the Rayleigh resolution limit of 0.81 (we note that the Rayleigh criterion which implies detectability at a signal to noise ratio of 5:1 is being used here only as a heuristic detectability threshold – at higher signal to noise ratios, smaller distances can be discerned from their smaller intensity dips). In Supplementary Note 1, we present analytical and numerical models to explore the interplay between sample thickness, gap size, and the effect of depth-blurring in determining the detection thresholds for MEP. These calculations show that MEP can detect gaps down to ~5 Å with a 20% contrast dip for a typical 2.4 nm FWHM Gaussian depth blur. Since van-der Waals distances in oxides are typically 4-5 Å, MEP is therefore capable of detecting such buried interfacial gaps. However, quantification of the precise gap extent is far more challenging as it requires knowing the bulk intensity level. This is retrievable from the MEP reconstruction only when both the membrane thickness is larger than twice the depth blur, and the gap is 5 Å or larger.

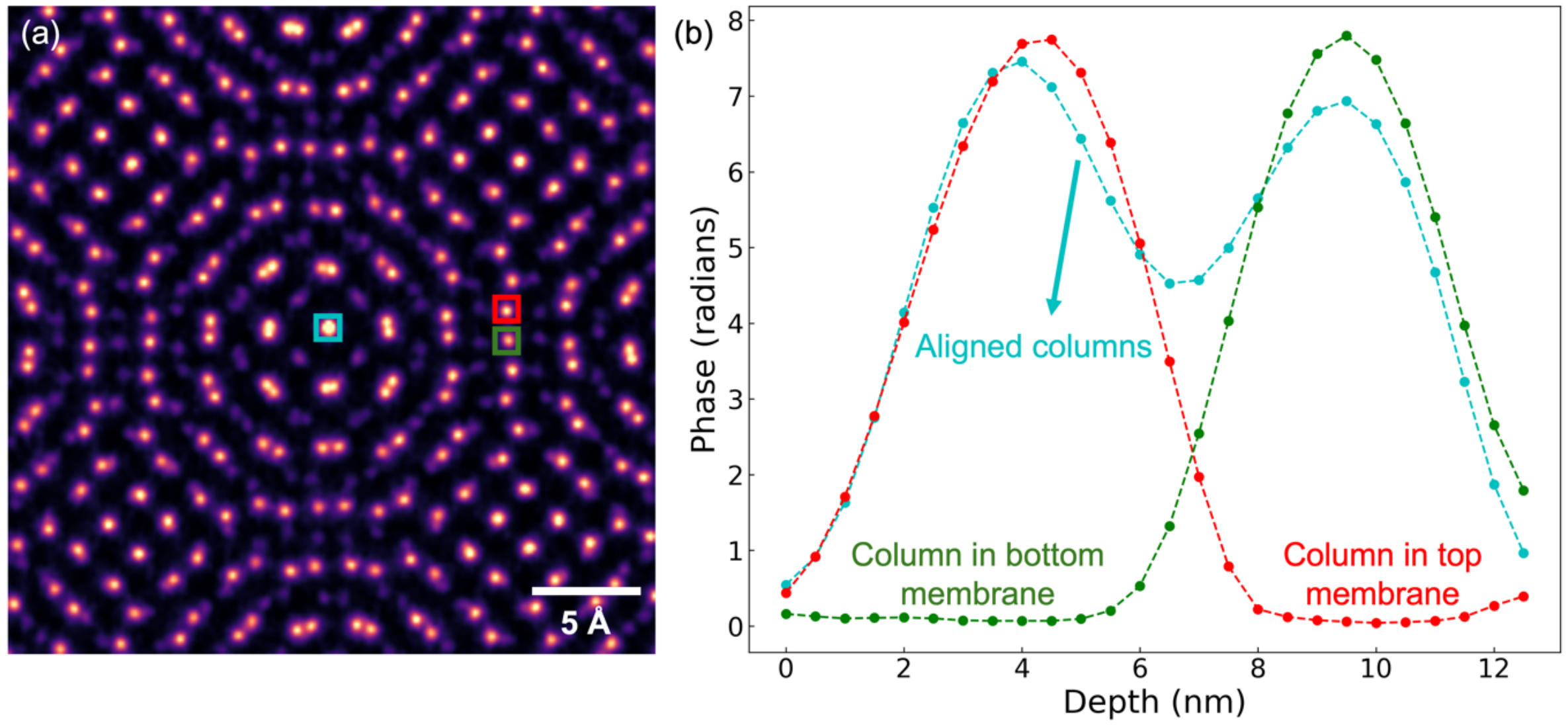

***Figure 4*. *Analysing depth resolution of MEP.*** *(a) Projected ptychographic reconstruction centered on an AA stacked region of the moiré pattern. Three positions are labelled, and the depth profiles of the respective atomic columns (summed over a 6x6 pixel area centered around the column) are shown in (b). The position labelled in red (green) has an atomic column present only in the top (bottom) layer as its depth profile decays rapidly in the other layer. The position labelled in cyan has columns present in both the top and bottom layers with the dip in the depth profile at the interface, indicating the presence of the gap between the two layers.*

Summing up depth profiles from the atomic columns confined to the top and bottom membranes (red and green profiles in Fig. 4(b)) results in a good match with the depth profile from the aligned columns (cyan profile) as shown in Supplementary Fig. 12, demonstrating the linearity of the imaging mode. A knife-edge analysis[33] on the red and green profiles in Fig. 4(b) gives an estimated depth resolution (difference in the depths between 90% and 10% of the maxima for the normalized phase range) of 2.31 nm and 2.46 nm respectively. We note that the knife-edge analysis will not yield meaningful results on the cyan profile in Fig. 4(b) as it does not show a plateaued minima on either side, a prerequisite for performing this analysis.

The depth profiles obtained with MEP were also directly compared with intensity profiles from cross-sectional imaging. Figure 5(a) shows the averaged line profile across the two membranes measured from the cross-sectional HAADF image shown in the inset (same as in Fig. 2(a)). The cross-sectional image shows evidence of finite surface roughness of the membranes, with a unit cell high step visible on the bottom surface. Hence, we calculate individual line profiles with unit-cell width from the cross-sectional image and overlay them in grey color with reduced opacity to show the spread in the intensity profile due to the surface roughness. We also note that the HAADF intensity is different for the two membranes due to variations in electron channeling arising from their relative crystallographic orientations. Since the MEP depth profile has a finite depth resolution, the HAADF cross-sectional line profile is blurred to match the minima-to-peak ratio of the MEP depth profile, with the comparison shown in Fig. 5(b). The blurring of the HAADF intensity profile required a Gaussian filter with a standard deviation of 1.34 nm or a full width at 80% maximum (FW80M) of 1.8 nm, with the latter being a direct estimate of the depth-resolution obtained with MEP. This estimate of the depth resolution surpasses the diffraction limit of 4.4 nm (depth of field given by $\frac{2\lambda}{\alpha^2}$, where semi-convergence angle ($\alpha$) =30 mrad, wavelength of 300 keV electrons ($\lambda$) =1.97 pm)[40] even without accounting for the additional chromatic blur, as scattered intensity in the dark field boosts information transfer of MEP to higher frequencies (upper limit given by $\frac{2\lambda}{\theta^2}$ = 0.8 nm, where collection angle ($\theta$) =69 mrad). Further details on how the depth resolution is influenced by dose, convergence and collection angles, and strength of the scattering species are discussed in Chen et al[27].

Comparison of the depth profiles in Fig. 5(b) indicates that MEP underestimates the overall sample extent in the depth direction by 11% as determined from peak-to-peak distance measurements. The surface roughness from the imaged area shown by means of an error band around the HAADF line profile in Fig. 5(b) is insufficient to explain this discrepancy. As the cross-sectional and plan-view images are taken at different sample regions, we also numerically estimate the surface roughness arising from step terraces of unit cell height (0.3905 nm) that could occur at the four different surfaces in the bilayer stack. Adding them in quadrature gives a net uncertainty of 0.79 nm (6%) for the sample thickness (double the peak-to-peak distance measured from the HAADF profile), also insufficient to explain the underestimation in sample thickness from the MEP depth profile.

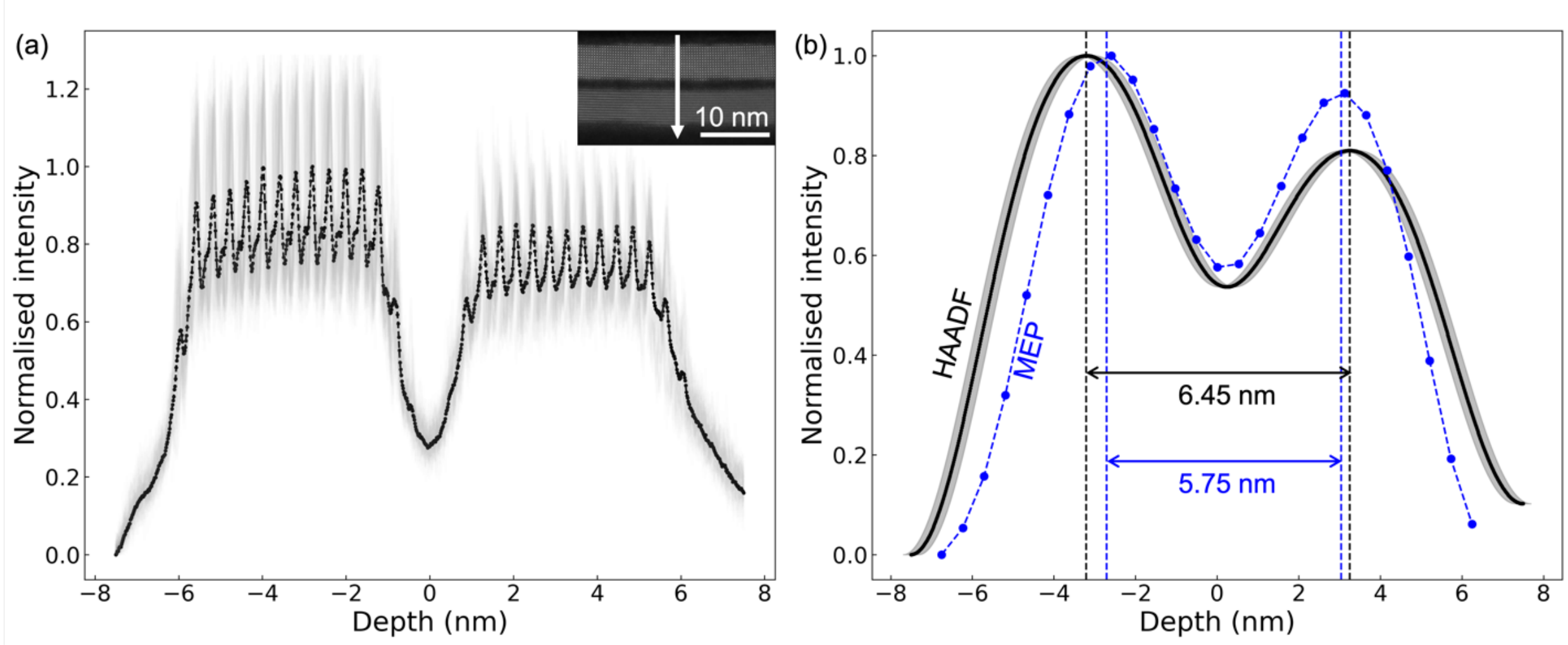

***Figure 5*. *Comparing accuracy of MEP with cross-sectional imaging.*** *(a) Averaged line profile through the cross-sectional HAADF image from Fig. 2(a) (shown in the inset), in bold, along with individual unit-cell-wide intensity profiles overlaid in grey with reduced opacity. (b) Gaussian-blurred HAADF cross-sectional profile plotted in black with an error band, showing the surface roughness, is compared with the MEP depth profile plotted in blue. The measured peak-to-peak distance from MEP is 11% lower than that measured from the HAADF cross-sectional line profile, indicating an underestimation of the sample thickness in MEP by 11%.*

Instead, this systematic error may result from the incomplete modelling of phonons in the forward scattering model currently used in ptychographic reconstruction codes where electrons are removed from the elastic channel using an absorptive potential but resulting inelastic Kikuchi patterns are not added to the simulation . Using multislice simulations (Supplementary Fig. 13), we show that phonons hinder effective electron channeling along atomic columns, shifting the dynamical diffraction effects to greater sample thicknesses. Using a purely elastic model of electron diffraction to match experimental diffraction patterns that contain the effect of electron-phonon scattering would result in an underestimation of sample thickness. For a sample thickness in the range of 10-15 nm, we measure this difference in thickness for dynamical diffraction effects with and without phonons to be ~12%, comparable with the measured discrepancy in experiment. However, MEP algorithms include an amplitude term in the object potential that attempts to account for phonon scattering through the absorptive model. In Supplementary Figs. 14 and 15, we show simulations that compare the purely elastic model, absorptive model, and frozen-phonon model. While the absorptive model and frozen-phonon model are well-matched at low scattering angles, the absorptive model is not a good approximation at large scattering angles. Hence, the reconstructed object thickness will depend on several experimental (probe convergence angle, outer collection angle, actual sample thickness, etc.) and computational factors (how the reconstruction algorithm weights low vs high scattering angles due to differences in scattered intensities, and any regularization on the amplitude term). Fully addressing this discrepancy requires the inclusion of a scheme to account for the effects of phonons in the MEP forward model beyond the simple absorptive model. While such models exist, they are quite computationally demanding[41–43].

We have shown that the presence of a moiré pattern in plan-view imaging with conventional STEM imaging methods is subject to artifacts. In the presence of a dead layer, a through-focal series with conventional STEM methods can be completely insensitive to the interfacial gap because of the multiple-scattering-dominated contrast mechanisms. We further demonstrated how these limitations can be addressed with multislice electron ptychography, enabling the detection of the interfacial dead layer. However, given the nanometer-scale depth-resolution of MEP, while the existence of a gap can be detected down to ~ 5Å, quantitative measurements of the interfacial gap can require cross-sectional images, which can routinely achieve sub-angstrom resolution. However, as we note in the discussion section below, cross-sectional HAADF images themselves are projections which will average over interfacial roughness through the thickness of the cross-section (~10-40 nm), making quantification of interfacial reconstructions challenging. We strongly caution against Fourier filtering of such images as it can introduce artificial atoms or atomic planes that misleadingly appear to bridge the gap as demonstrated in Supplementary Fig. 16. In such cases, 3D imaging becomes necessary to reveal details that are concealed in projection. Therefore, a combination of cross-sectional imaging with HAADF or MEP, and plan-view depth-sectioning with MEP is needed to achieve a comprehensive characterization of interfacial proximity and potential structural reconstructions in these moiré structures.

## 5. Hybrid interface between a 2D material and oxide membrane

Although our discussion above focused on a bilayer stack of oxide membranes, similar challenges also exist for interfaces between as well as 2D materials and oxide membranes[44–46]. An example of such a mixed interface is illustrated in Fig. 6 for a $MoS_2$ (monolayer) / $SrTiO_3$ / $MoS_2$ (monolayer) heterostructure. Figure 6(a) shows a large field-of-view cross-sectional HAADF image of the heterostructure, revealing a visible gap between the layers, interspersed with small patches of atomic-scale proximity (labelled with yellow arrows). Using electron energy loss spectroscopy (EELS), we show that both top and bottom interfaces are clean with no carbon contamination layer (typically present at such interfaces if the transfer is not clean) between the 2D material and the oxide membrane, with details of the analysis shown in Supplementary Fig. 17. Magnified images of the top and bottom interfaces are shown in Fig. 6 (b, c) respectively and used to calculate an average projected interlayer distance (Mo plane – terminating layer of $SrTiO_3$) of $5.59 \pm 0.87$ Å and $6.37 \pm 1.50$ Å for the top and bottom interfaces respectively. More images of both interfaces and interlayer measurements are presented in Supplementary Figs. 18, 19. Accounting for the Mo-S interplanar distance of 1.62 Å, this measured interlayer distance (~ 4 Å) is still slightly larger than typical van-der Waals distances (2.9 Å for bulk $MoS_2$). Although the carbon contamination layer is eliminated, an overall interfacial gap persists as the $MoS_2$ monolayer cannot conform perfectly to the surface of the $SrTiO_3$ membrane. Such roughness of the $SrTiO_3$ substrate has been noted to affect electrical measurements of supported 2D materials [47,48]. When feasible, atom-by-atom growth of the 2D material on an oxide substrate offers an alternative route to film transfer to creating conformal structures[49,50].

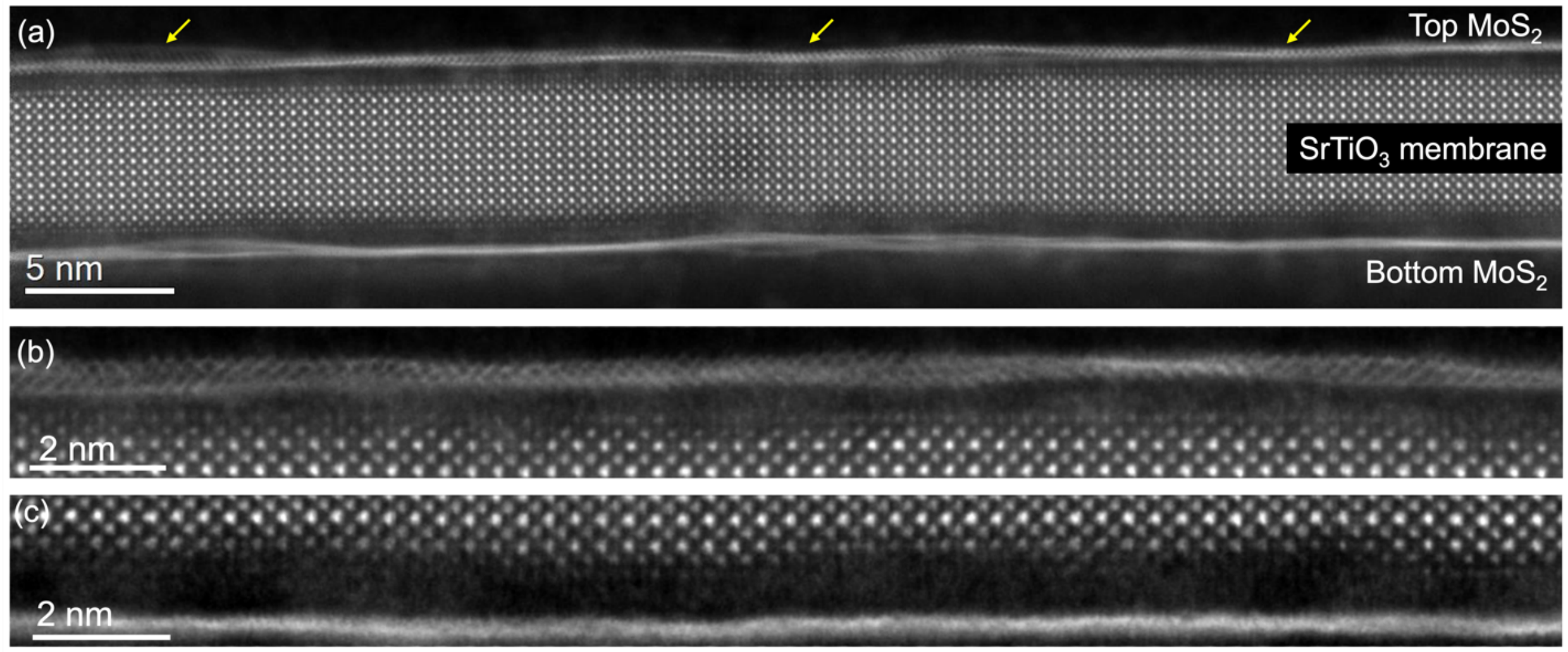

***Figure 6*. Hybrid 2D-3D interfaces.** *(a) Cross-sectional HAADF image of $MoS_2$ (monolayer) / $SrTiO_3$ / $MoS_2$ (monolayer) heterostructure. Patches where the 2D material and oxide membrane are atomically close are marked with yellow arrows. (b, c) Magnified images of the top and bottom interfaces. $MoS_2$ does not conform fully to the $SrTiO_3$ membrane due to the surface roughness of the oxide membrane.*

## 6. Discussion

As motivated above, even a contamination-free interface does not guarantee a conformal interface between stacked membranes as even atomically clean surfaces inevitably contain steps, terraces and islands whose extremities can limit the areal fraction over which true atomic proximity is achieved. The extent of interfacial conformation is governed by the competition between adhesion and bending energies. Following previous work[51,52], we estimate a critical bending radius of 2.8 ± 0.7 nm for monolayer $MoS_2$ on $SrTiO_3$. This means that the $MoS_2$ film cannot conform to roughness details of the substrate that have radii of curvature below 3-4 nm, which is consistent with the length scale of bends seen in Fig. 6. The bending stiffness (B) of thin $SrTiO_3$ membranes varies with the membrane thickness (t) as $B \propto t^3$, so the critical radius for a stacked heterostructure between $SrTiO_3$ membranes depends strongly on the membrane thickness. For a 5-10 nm thick $SrTiO_3$ membrane, the critical radius for van der Waals forces is 70-200 nm, and we would expect $SrTiO_3$ membranes to have trouble conforming to features smaller than this length scale for van der Waals forces. For direct chemical bonds where adhesion energies are roughly 100x larger, the critical bending radius is 7-20 nm. Details of the calculation are given in Supplementary Note 2.

Given the large critical bending radii of $SrTiO_3$ membranes, stacking two such membranes is unlikely to generate a conformal interface without substantial thermal processing. As noted above, even perfectly clean surfaces exhibit steps, terraces and islands, and because the surface roughness of two membranes is completely uncorrelated, we would expect stacking to typically yield atomic contact only at isolated extremities. Heating to temperatures above the step-flow regime can promote atomic diffusion, enabling interfacial rearrangement toward a more conformal interface. For $SrTiO_3$ this step-flow temperature is around 800 °C[53] and annealing steps at higher temperatures for thin membranes on bulk substrates have indeed produced a conformal interface[14].

Given the potential for imaging artifacts noted here, we recommend a protocol for robust characterization of these stacked membranes. For plan-view imaging which does not require the preparation of a cross-sectional foil, we recommend using MEP as a first step to characterize the buried interface – either to detect a possible gap or for quantifying the moiré strain field if there is no gap. In this study, we have experimentally demonstrated the capability of MEP to detect buried interfacial gaps when conventional through-focal imaging methods fail. Moreover, our simulations (Supplementary Note 1 and Supplementary Fig. 20) show that gaps down to the 1-2 unit-cell level or close to typical van-der Waals distances of 4-5 Å can be detected with the typical depth blurring of a good MEP setup of 2-3 nm. However, quantification of this gap extent is challenging and requires knowledge of the bulk intensity, which is accessible from the MEP reconstruction only when each membrane is thicker than twice the MEP depth blur and if the gap larger than ~ 4-5Å.

When further characterization of the interface or chemical mapping is required, cross-sectional imaging is also recommended – EELS in projection will suffer similar (or even worse) multiple scattering artifacts to those we showed for HAADF[54]. However, care must be taken in the interpretation of cross-sectional images, as cross-sectional images are still projections through 10-40 nm thick TEM samples. When atomic registry occurs only in localized patches, cross-sectional imaging can give the illusion of atomically-conformal interface as illustrated in Fig. 7 (a-d). In such cases, the appearance of a darker band at the interface in either the image or the spectral map is a good indication that the true contact area is substantially diminished. As the foil thickness is typically much larger than the length scale of surface roughness, directly reading off the interface separation from cross-sectional images will also result in an underestimation of gap width. For a perfectly coherent interface as shown in Fig. 7(e), there would be no intensity dip at the interface as shown in Fig. 7(f), although the intensity of the image on either side of the interface may be different due differences in electron channeling along different crystal orientations.

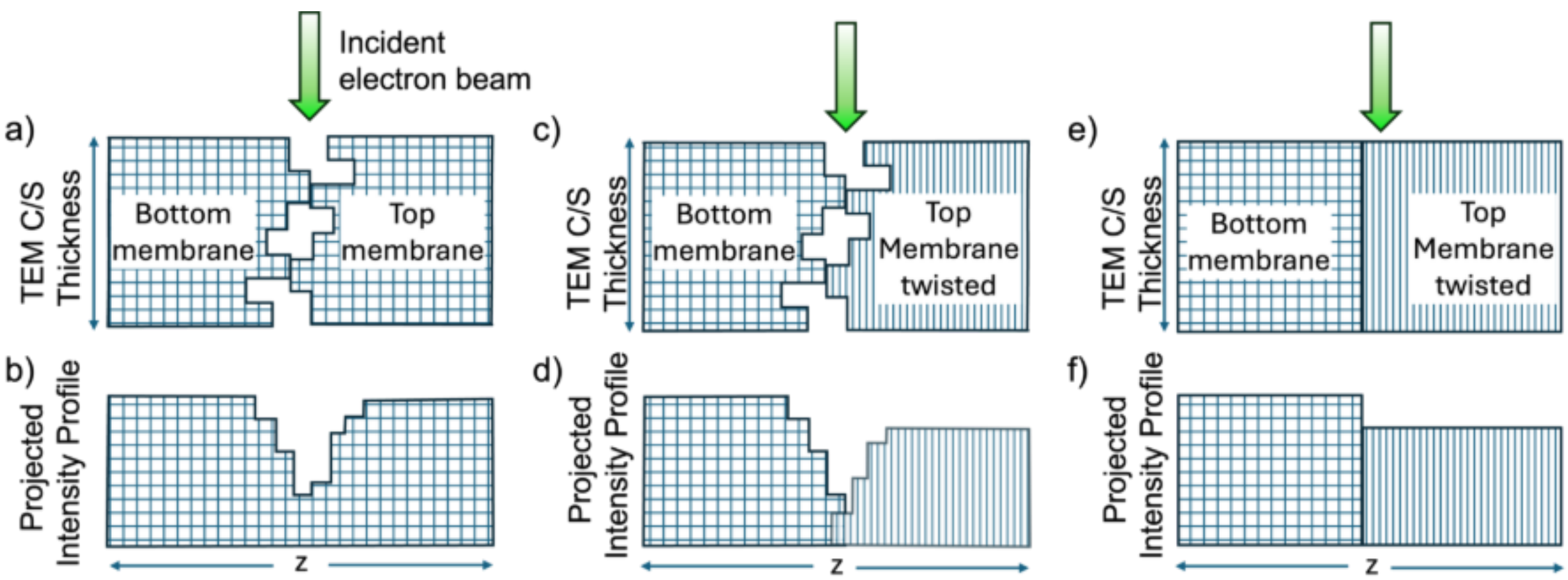

***Figure** 7. **Effect of surface roughness on cross-sectional (C/S) images.** The C/S is prepared by FIB cut into the membrane and removing a sliced coupon that is laid on its side for TEM imaging. a) shows two membranes with rough, uncorrelated surfaces that only contact at a few extremities, and b) shows the projected mass thickness imaged by conventional TEM or STEM. Note the reduction in contrast at the interface, but the overlapping (but unconnected lattices) creates the illusion of a continuous lattice. When the one membrane is rotated c) shows that lattice fringes in the plane of rotation are preserved but those perpendicular to the rotation are lost. In projection (d), the overlapping appears to blend into each other, again creating the illusion of a continuous lattice, but the reduced intensity again is a clear marker that the contact area is substantially reduced. If the interface were atomically smooth (e) then there would be no intensity reduction (f) in the intensity profile at the interface. In other words, a dark band at the interface in either an HAADF or EDX profile is good indicator of an imperfect interface.*

## 7. Conclusion

As detailed above, there remain significant challenges and opportunities for improving the quality of interfaces in bilayer membranes. Unlike van der Waals materials which have a stable surface termination, oxide membranes often form contamination layers at their surfaces, especially on exposure to environmental factors during the liftoff or transfer processes. The surface roughness of the oxide membranes also needs to be carefully controlled to prevent interfacial coupling from occurring only at isolated patches. Meticulous sample preparation including, but not limited to, careful surface treatments, cryogenic transfer protocols, and high-temperature annealing steps, can produce atomically sharp interfaces in related material systems as demonstrated in recent reports with a twisted stack of layered high-temperature cuprate superconductors[55] (interface between exfoliated flakes of a van der Waals oxide) and an epitaxially incompatible symmetry-forbidden interface between $SrTiO_3$ and sapphire[14] (membrane-single crystal interface). As interlayer electronic coupling in a stacked bilayer is reliant on close proximity of the two layers and is an inherent assumption in theoretical modeling of moiré materials or hybrid 2D-3D heterostructures, optimizing the interface will be key in unlocking the new physics that they may hold.

**Acknowledgements**
H.K., K.J.C., Y.-T.S., J.-H.K., C.L., J.P., H.Y.H., and D.A.M. acknowledge funding from the Department of Defense, Air Force Office of Scientific Research under award FA9550-18-1-0480. X.W., Y.L., and R.X. acknowledge support from the US Department of Energy, Office of Basic Energy Sciences, Division of Materials Sciences and Engineering, under Contract No. DE-AC02-76SF00515. C.-H.L. would like to thank the support from the Eric and Wendy Schmidt AI in Science Postdoctoral Fellowship, a program of Schmidt Sciences, LLC. R.X. acknowledges the support from the National Science Foundation (NSF) under award No. DMR-2442399. This work made use of the electron microscopy facility of the Platform for the Accelerated Realization, Analysis, and Discovery of Interface Materials (PARADIM), which is supported by the National Science Foundation under Cooperative Agreement No. DMR-2039380 and Cornell Center for Materials Research shared instrumentation facility with Helios FIB supported by NSF (DMR-1539918). The authors thank John Grazul, Mariena Silvestry Ramos, Steven Zeltmann, Phil Carubia, and Malcolm Thomas for technical support and maintenance of the electron microscopy facilities. The authors also thank Dr. Varun Harbola, Prof. Darrell Schlom, and Prof. Kyle Shen for insightful discussions.

## Supplementary Information

## Methods

### $SrTiO_3$ thin-film growth

15 u.c. $SrTiO_3$ / 20 u.c. $Sr_2CaAl_2O_6$ heterostructures were synthesized on 5 × 5 mm$^2$ (001)-oriented single-crystalline $SrTiO_3$ substrates by pulsed laser deposition (KrF, $\lambda$ = 248 nm). $SrTiO_3$ substrates were *in situ* pre-annealed at 930 °C under oxygen partial pressure $P_{O_2} = 5 \times 10^{-6}$ Torr for 30 minutes to obtain a step-and-terrace surface. The sacrificial layer $Sr_2CaAl_2O_6$ was grown at 700 °C with $P_{O_2} = 5 \times 10^{-6}$ Torr, laser fluence $F$ = 3.12 J cm$^{-2}$ (spot size = 2.20 mm$^2$) and laser repetition frequency $f$ = 1 Hz. $SrTiO_3$ layer was grown at 700 °C with $P_{O_2} = 5 \times 10^{-6}$ Torr, $F$ = 0.89 J cm$^{-2}$ (spot size = 3.20 mm$^2$), and $f$ = 1 Hz. A single-crystal $SrTiO_3$ target was used for the $SrTiO_3$ growth, while polycrystalline ceramic target was used for $Sr_2CaAl_2O_6$ growth.

### Twisted $SrTiO_3$ membrane fabrication

The $SrTiO_3$ thin films were spin-coated with A8 950 polymethyl methacrylate (PMMA) at 3,500 r.p.m. for 60 seconds, with a ramping rate of 1,000 r.p.m. s$^{-1}$, achieving a thickness of approximately 1 μm. Following spin-coating, the films were cured at 135 °C for 7.5 minutes. The samples were immersed in deionized water for 6 hours to dissolve the sacrificial layer. The $SrTiO_3$ membrane were then transferred onto 200 nm-thick perforated $SiN_x$ membranes with circular holes in diameter of 2 μm, supported by 200 μm Si frame. For improved bonding, the samples were heated at 110 °C for 10 min before being dipped into acetone and isopropanol subsequently for 10 min to dissolve the PMMA. The freestanding $SrTiO_3$ membranes on TEM grids were annealed in a tube furnace under ambient atmosphere: 700 °C for 1 hour for sample 1 and 900 °C, 3 hours for Sample 2.

To minimize air exposure and maintain a clean interface, the second-layer $SrTiO_3$ membrane was immediately transferred onto the annealed first-layer $SrTiO_3$ with the desired twist angle. After the second transfer, the PMMA removal and anneal conditions were repeated for second $SrTiO_3$ layer to obtain a twisted $SrTiO_3$ device.

### $MoS_2$ thin-film growth

Monolayer $MoS_2$ films were synthesized by metal organic chemical vapor deposition. Molybdenum hexacarbonyl (MHC) and diethyl sulfide (DES) are chosen as chemical precursors for Mo and S, respectively. They are introduced to a growth chamber in gas phase with $H_2$ and $N_2$. We use a total pressure of ~10 Torr and growth temperature of ~ 600 °C. The flow rates of precursors are 5 sccm for MHC, 0.12 sccm for DES, 1 sccm for $H_2$, and 1500 sccm for $N_2$, which were regulated by mass flow controllers.

### $MoS_2$ membrane fabrication

The steps for separation of van der Waals materials from their substrates were:

- Spin coating of an adhesive polymer layer of PMMA (Poly-methyl methacrylate, 495 K, 4% diluted in anisole) for 90 seconds at 4000 rpm on an as-grown monolayer $MoS_2$ film sitting on its growth substrate ($SiO_2$/Si).
- Baking 10 mins at 180°C using a hot plate, followed by attaching a thermal release tape (TRT) manufactured by Nitto-Denko.
- TRT/PMMA/$MoS_2$ is then separated from the $SiO_2$/Si substrate via mechanical peeling without the use of any chemicals or etchants, which keeps the bottom surface of $MoS_2$ clean.

**Fabrication of 2D material – oxide interfaces**

Fabrication of vdw/non-vdW heterostructures was performed as follows. First, fabrication of $MoS_2$/$SrTiO_3$ heterostructure (vdW on top of non-vdW) is done by vacuum stacking, mounting TRT/PMMA/$MoS_2$ on the top stage holder and putting another as-grown $SrTiO_3$ film on the bottom stage of the vacuum box. The chamber was evacuated to less than 200 mTorr and the bottom stage heated to 150°C. The top holder was lowered to make a contact between $MoS_2$ and the $SrTiO_3$/$Sr_2CaAl_2O_6$/ $SrTiO_3$ heterostructure using the z-motion linear vacuum feedthrough, keeping them in contact for 10 mins. Then the top holder was lifted with the stacked sample. After these steps, TRT/PMMA/$MoS_2$ is attached to the $SrTiO_3$/$Sr_2CaAl_2O_6$/$SrTiO_3$ heterostructure, forming a pristine $MoS_2$/$SrTiO_3$ interface. The sample is then taken out and placed in deionized water at room temperature until the sacrificial $Sr_2CaAl_2O_6$ layer is fully dissolved. Next, to make a heterostructure of $MoS_2$/$SrTiO_3$/$MoS_2$ (non-vdW on top of vdW), the TRT/PMMA/$MoS_2$/$SrTiO_3$ was mounted on $MoS_2$/$SiO_2$/Si and stacked in vacuum system, and then separation process of TRT/PMMA/$MoS_2$/$SrTiO_3$/$MoS_2$ from $SiO_2$/Si was repeated as explained above. TRT is easily removed by hot plate annealing and PMMA is removed by acetone and isopropanol to generate $MoS_2$/ $SrTiO_3$/$MoS_2$ hybrid heterostructures.

**Scanning transmission electron microscopy (STEM)**

The cross-sectional samples were prepared using a Thermo Fisher Helios G4 UX focused ion beam using the standard lift-out method. The plan-view imaging was done with the sample on a TEM grid with 2 um holes. All STEM data on the twisted bilayer oxide sample was collected using an aberration corrected Thermo Fisher Spectra X-CFEG STEM, operated at 300 kV, probe semi-convergence angle of 30 mrad and probe current of 60 pA. The collection angle for HAADF images was 62-200 mrad. DPC data was acquired on the Panther segmented detector with a collection angle of 15-55 mrad. The EDX data was acquired on a Dual-X detector with a dispersion of 5 eV and a frame time of 5.2 seconds, with 439 frames acquired in total. The through-focal series was also acquired under the same conditions with defocus steps of 1 nm. The defocus was first adjusted to place the probe focus outside the sample and subsequently stepped through the sample. Data on the $MoS_2$/$SrTiO_3$/$MoS_2$ heterostructure was acquired with an aberration-corrected FEI Titan Themis X-FEG STEM (300 keV beam energy and probe semi-convergence angle of 21.4 mrad). The EEL spectra were acquired using a 965 GIF Quantum ER spectrometer and a Gatan UltraScan scintillator-based detector.

**Multislice electron ptychography (MEP)**

4D-STEM datasets are acquired using the EMPAD-G2 detector[1] operated at 10 kHz and reconstructed with PtyRAD[2]. Major experimental and reconstruction parameters are organized in SI Table 1. 4D-STEM data and input parameter file will be made publicly available for reproducibility. MEP object slices are further aligned using a registration algorithm[3] to account for tilt differences between the top and bottom membranes. The 3D volume rendering of the MEP stack of slices and HAADF through-focal series is done using tomviz[4].

**Supplementary Table 1: Experimental and reconstruction parameters for MEP**

| Experimental parameters | Value |
|---|---|
| **Sample thickness** | 15 nm |
| **Twist angles** | Sample 1: 9°; Sample 2: 45° |
| **Acceleration voltage** | 300 kV |
| **Probe convergence semi-angle (α)** | 30 mrad |
| **Probe defocus** | 6 nm (overfocus) |
| **Detector type** | EMPAD G2 (128 x 128 pixels) |
| **Camera length** | 300 mm |
| **Maximum collection angle (θ)** | 69 mrad, or 3.52 $Å^{-1}$ |
| **k-space sampling** | 1.08 mrad, or 0.055 $Å^{-1}$ |
| **Scan step size** | 0.415 Å |
| **Total scan size** | 10.62 nm x 10.62 nm (256 x 256 pixels) |
| **Dwell time** | 100 μs |
| **Dose** | $2 \times 10^5$ electrons/$Å^2$ |
| **Reconstruction parameters** | **Value** |
| **Number of object slices** | 30 |
| **Slice thickness** | 5 Å |
| **Depth regularization** | Real space axial Gaussian blur (obj_zblur = 1) |
| **Number of probe modes** | 6 |
| **Learning rate (Object)** | 5.0e-4 |
| **Learning rate (Probe)** | 1.0e-4 |
| **Learning rate (Position)** | 5.0e-4 |
| **Optimizer** | Adam |
| **Noise model** | Amplitude model with Gaussian noise (loss_single: weight = 1, dp_pow = 0.5) |
| **Object phase positivity** | On |
| **Object phase L1 sparsity** | On (loss_sparse: weight = 0.1, ln_order = 1) |
| **Diffraction pattern blur** | 0.5 pixels |
| **Batch size** | 32 |
| **Iterations** | 300 |

**Schematic for parallax effect**

Outlines for the duck and cloud used in the parallax schematic shown in Fig. 1 were first hand-drawn, followed by digitization using the OpenCV python package. Sr atoms were then placed along the outlines to create the objects used in the simulation. The multislice calculations for simulation of the diffraction patterns were carried out using the abTEM[5] simulation suite and used a beam energy of 300 keV and a probe semi-convergence angle of 30 mrad. The probe is overfocused 50 nm and 250 nm with respect to the duck and cloud respectively.

## Supplementary Figures

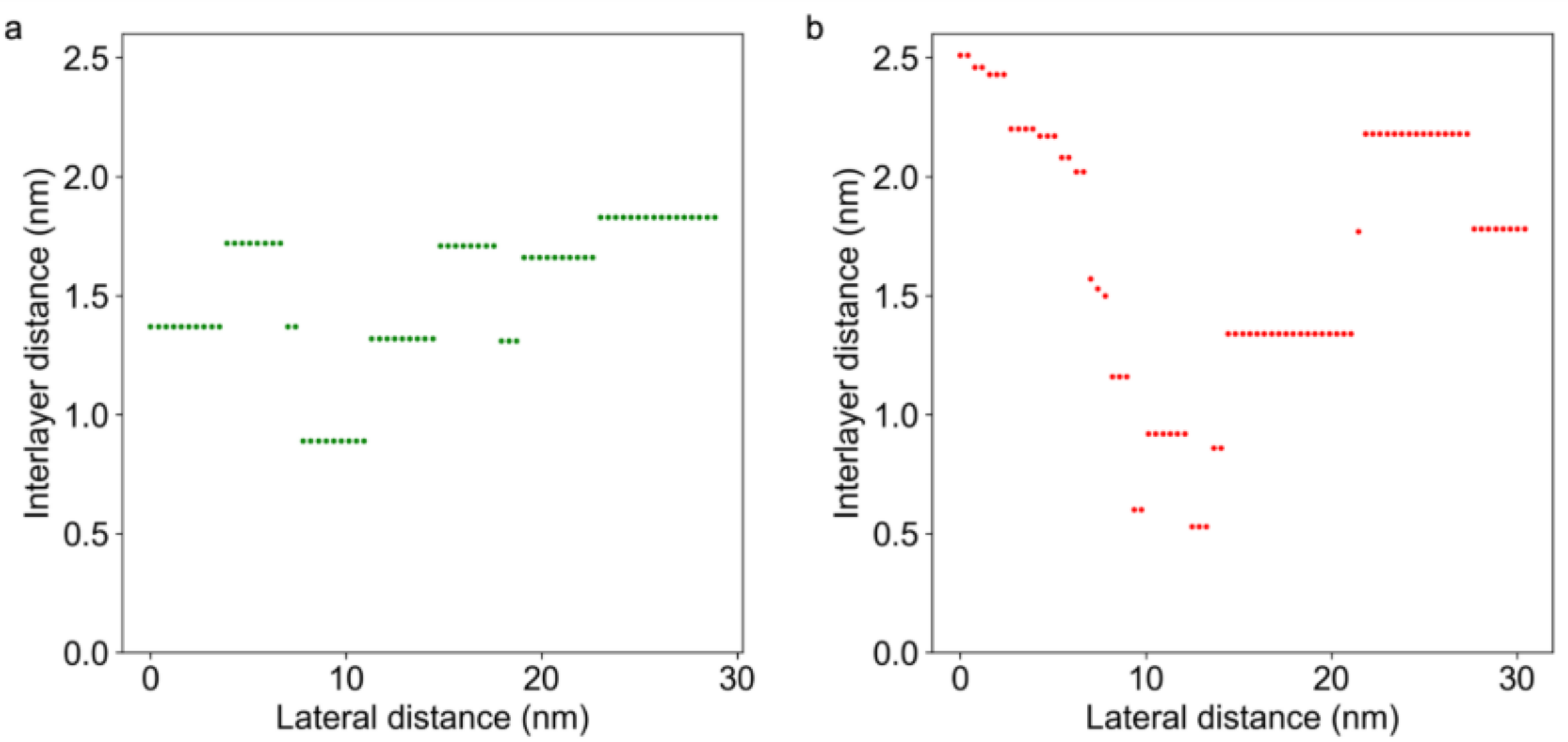

**Supplementary Fig.** 1 | Measured values of projected interlayer distances between the two oxide membranes for (a) Sample 1 and (b) Sample 2. The measurement is done from the same images shown in Fig. 2.

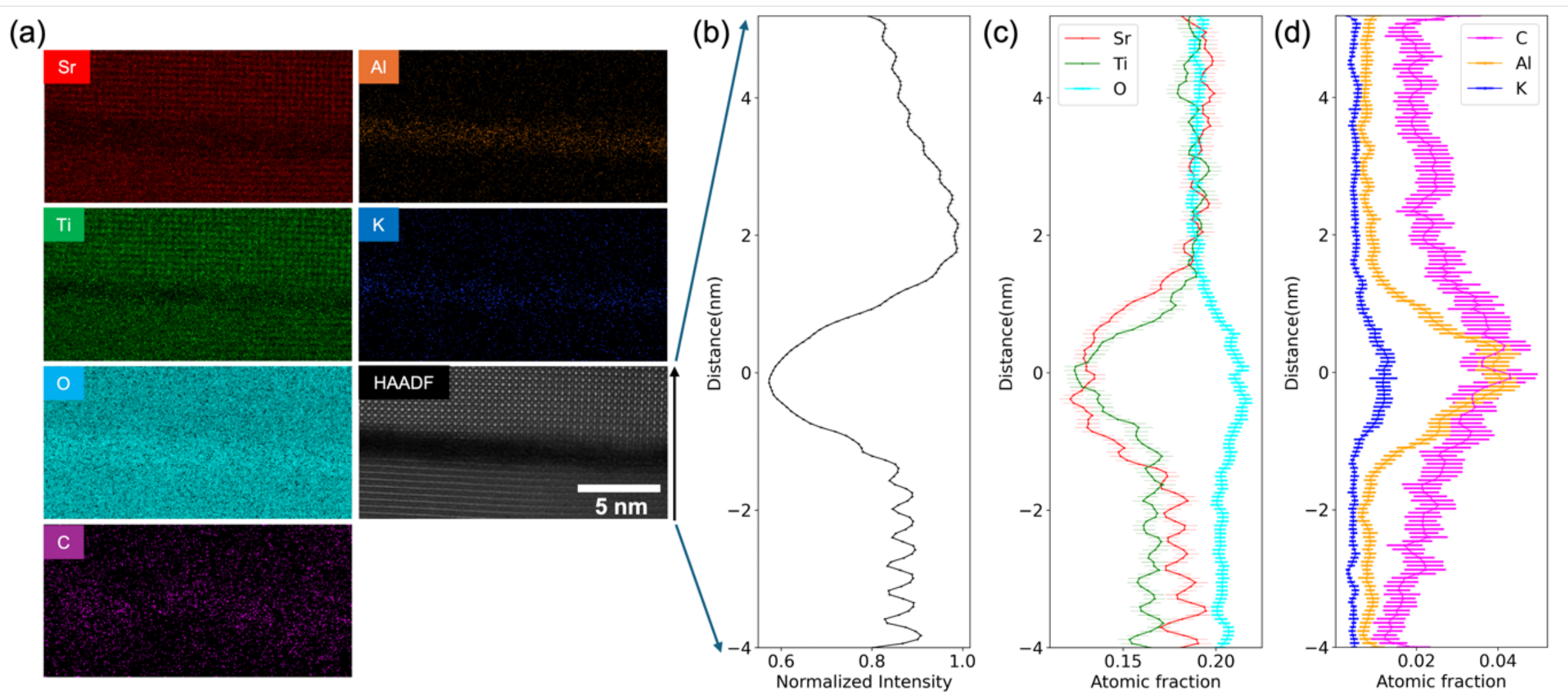

**Supplementary Fig. 2** | Energy dispersive X-ray (EDX) spectroscopy on Sample 1 - cross-sectional sample of the twisted $SrTiO_3$ bilayer. (a) Elemental maps of the different elements detected in the EDX spectrum. (b-d) Line profile (averaged parallel to the dead layer) of the (b) HAADF image intensity, (c) atomic fraction of Sr, Ti, O, and (d) atomic fraction of C, Al, K. We note that the atomic fraction of O is rescaled by a factor of 1/3. The line profiles show the presence of higher amounts of C, O, Al and K in the gap region, indicating the presence of organic materials as well as trace amounts of remanent oxides from the sacrificial layer.

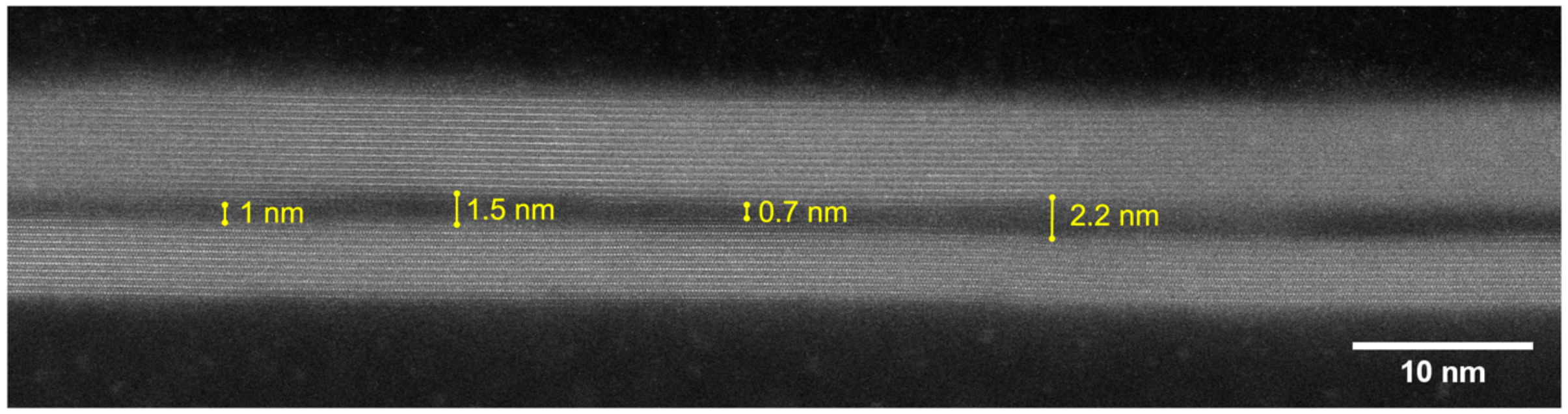

**Supplementary Fig. 3** | Cross-sectional HAADF image of Sample 2 (bilayer $SrTiO_3$ membrane sample) with large variations in the extent of the interfacial gap due to the surface roughness of the top and bottom membranes.

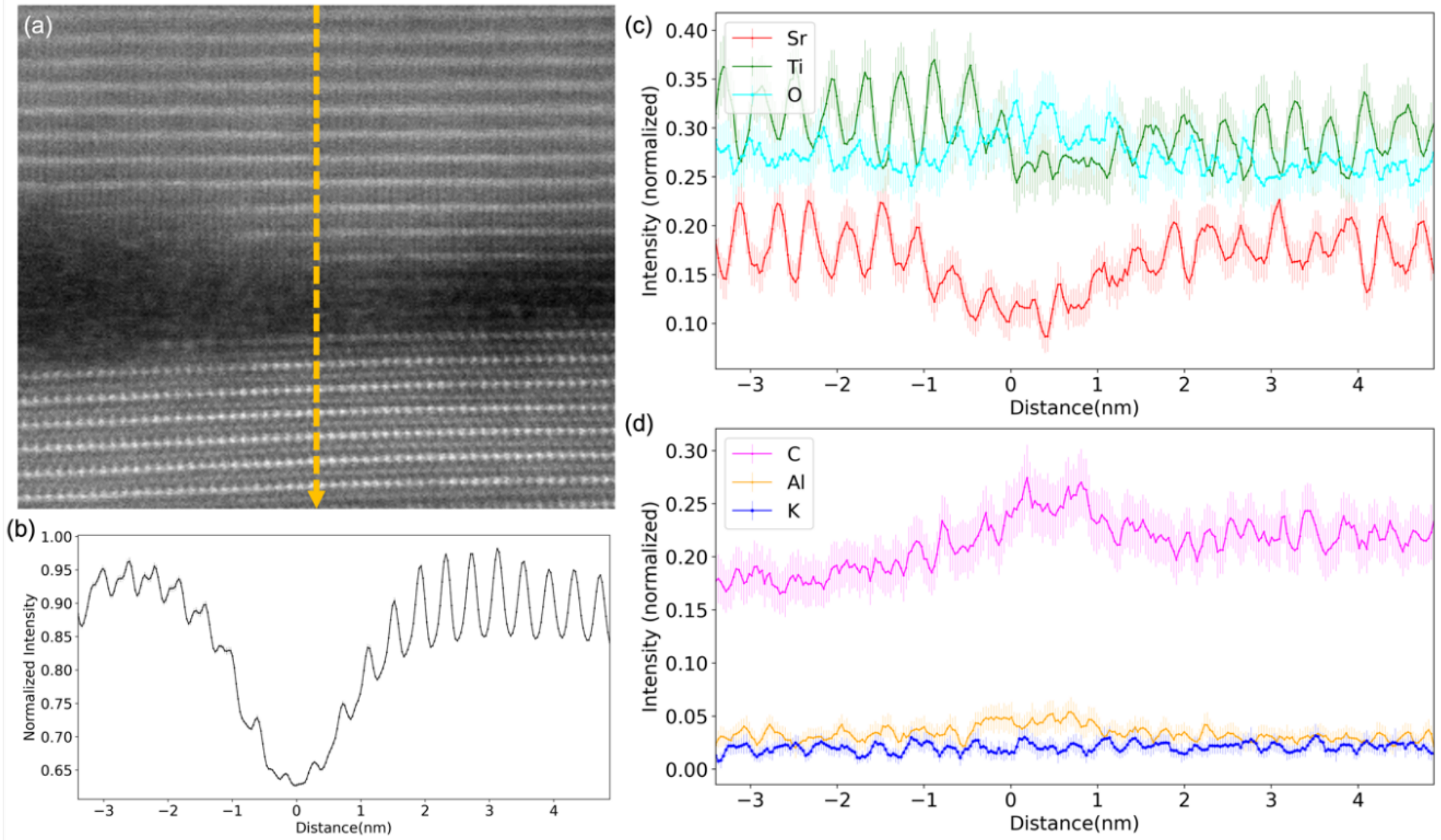

**Supplementary Fig. 4** | Energy dispersive X-ray (EDX) spectroscopy on Sample 2. (a) Cross-sectional HAADF image of the bilayer. (b) Line profile of the HAADF image intensity along the yellow dotted line in (a) and averaged in the perpendicular direction. Normalized intensity of (c) Sr (L-lines), Ti (K-lines), O (K-lines), and (d) C, Al, K (K-lines for all). The normalization is such that the total counts for each data point for all elements sums up to 1. The line profiles show higher amounts of C and O in the gap region. 2D elemental maps and atomic fractions are not presented as the counts were too low for any quantification.

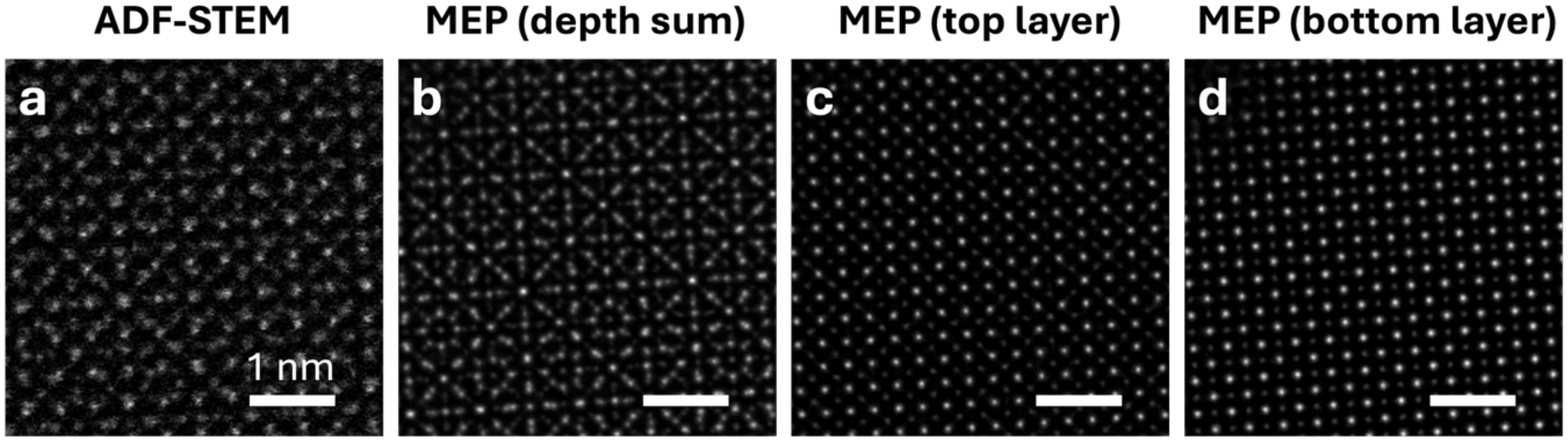

**Supplementary Fig. 5** | Plan-view characterization of twisted bilayer SrTiO3 with (a) ADF-STEM and (b-d) multislice electron ptychography (MEP). (b) Depth-sum image of MEP (c) Reconstructed slice at depth = 3 nm inside the top layer (d) Reconstructed slice at depth = 3 nm inside the bottom layer. While both ADF-STEM and MEP show clear moiré pattern with a twist angle around 45 degrees, MEP provides unique capability of resolving individual membranes with better spatial resolution. All scale bars are the same.

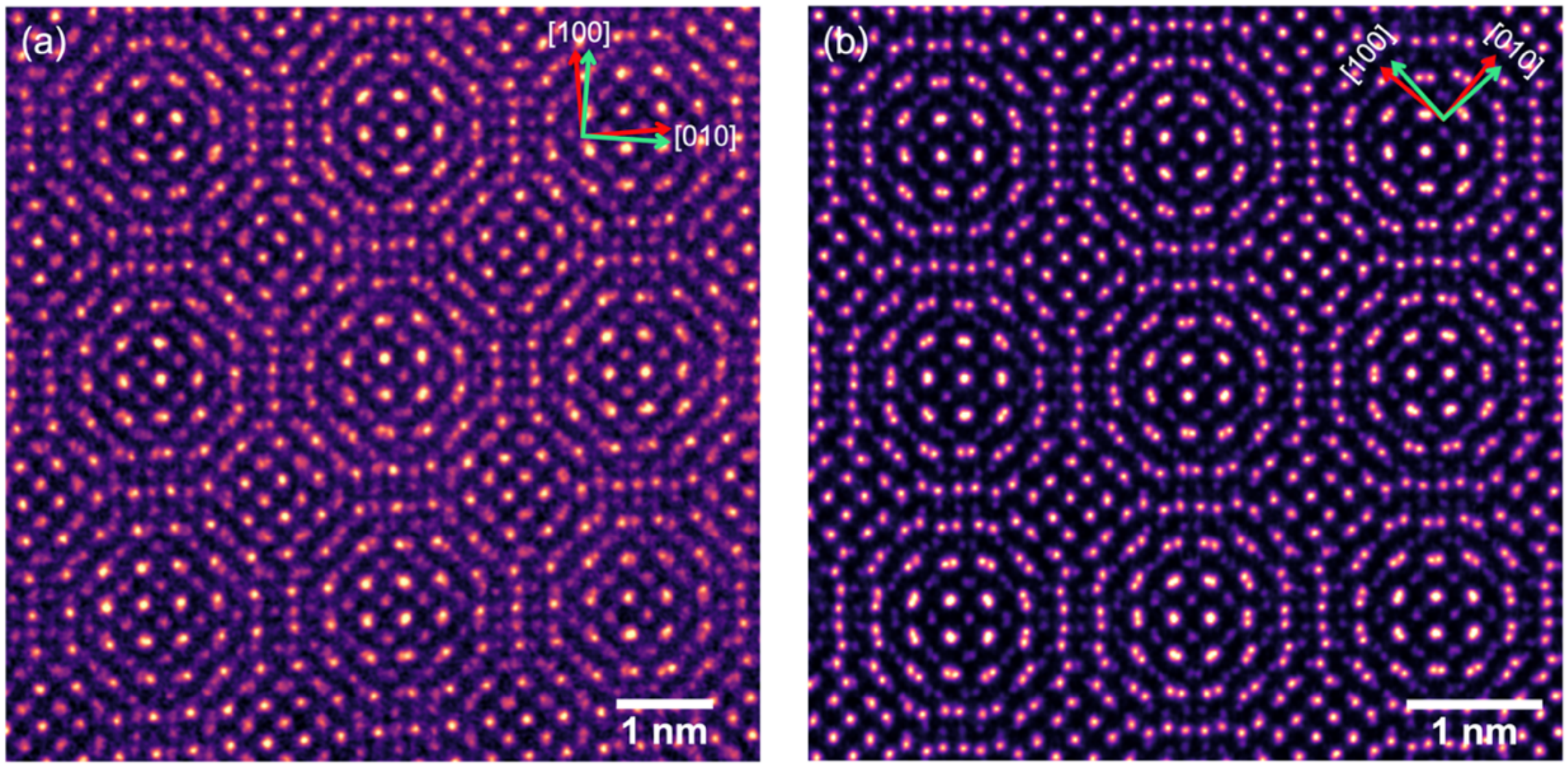

**Supplementary Fig. 6** | Projected moiré patterns of Sample 1 from (a) HAADF, and (b) multislice ptychography that look visually identical, although they cover different spatial extents. The spatial extent of the field of view in the HAADF image is roughly $\sqrt{2}$ times larger and rotated by 45° with respect to the ptychographic image. The similarity in the patterns after the above geometrical transformations results from the different scaling of the image contrast with atomic number and is illustrated in detail in the following Supplementary figures.

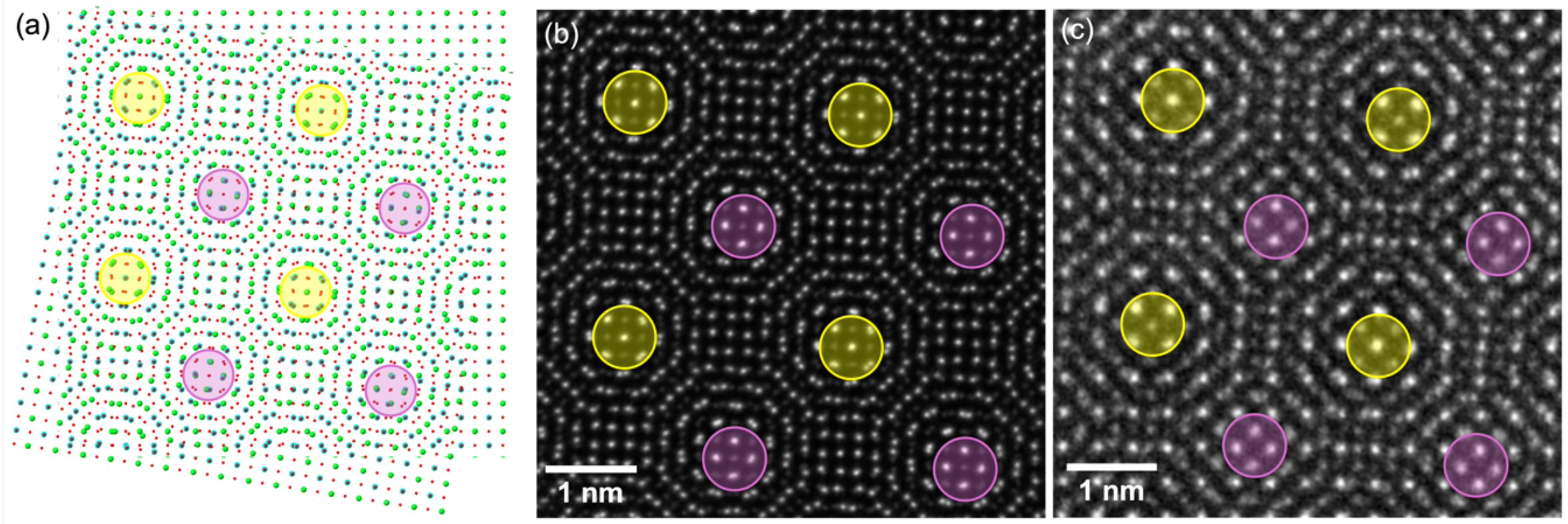

**Supplementary Fig. 7** | (a) Schematic of a moire pattern created by a 9° twist angle between 2 $SrTiO_3$ layers with AA (AB) stacked regions marked with yellow (plum) color. (b) Projected multislice ptychographic reconstruction of the twisted bilayer $SrTiO_3$ sample (Sample 1) displaying a similar pattern. Due to the similar contrast of the Sr columns and Ti-O columns in multislice ptychography, the AA stacked regions centered at the A-site or the B-site look visually similar. (c) Projected HAADF image of the twisted bilayer $SrTiO_3$ sample (Sample 1) – the moire pattern looks visually different from the schematic because the oxygen atoms are not present. The AA stacked regions centered at the A-site and B-site can be distinguished due to the stronger contrast of the Sr column with respect to the Ti column.

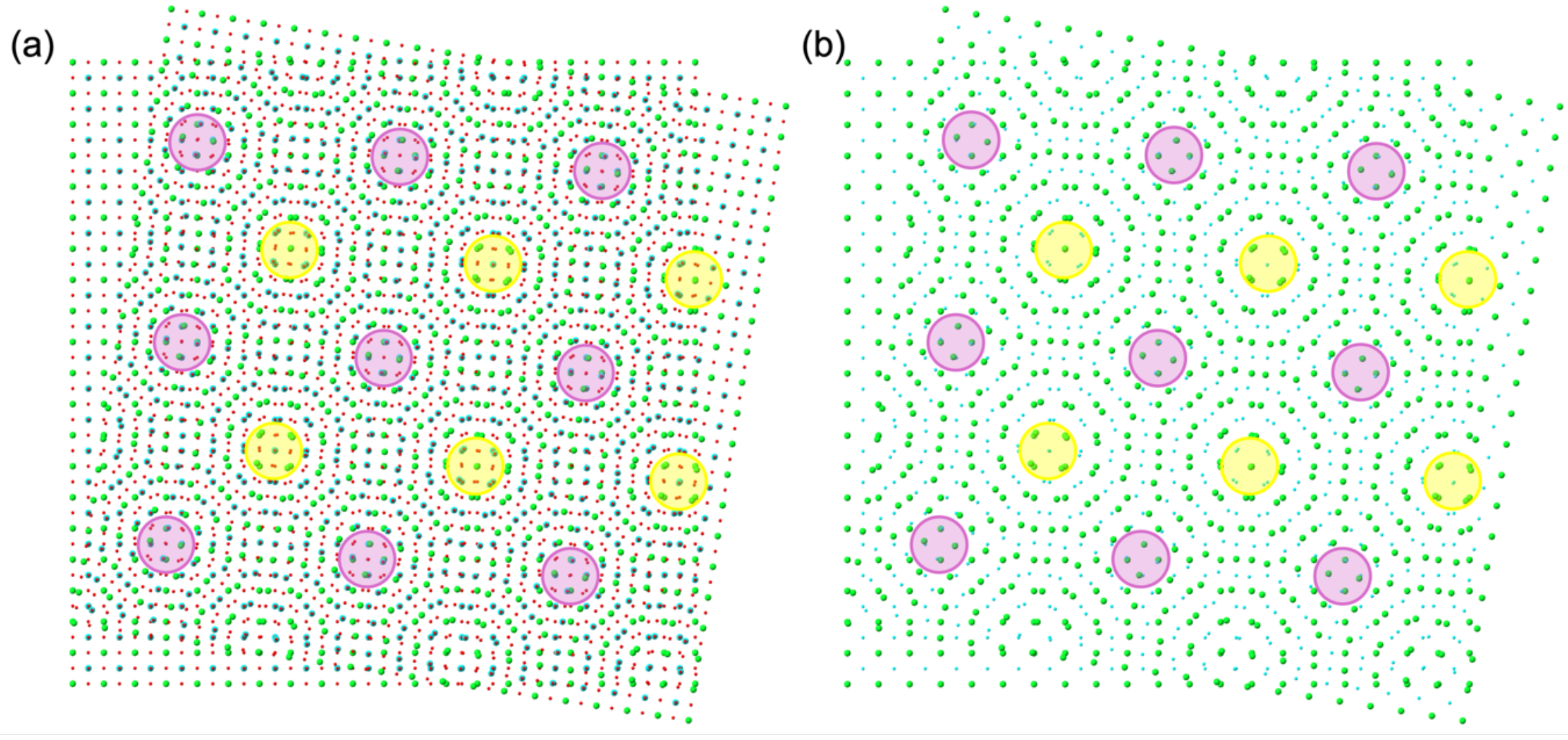

**Supplementary Fig. 8** | Overlay of projected crystal structures of $SrTiO_3$ with a relative twist angle of 9° to mimic the moiré pattern in (a) multislice ptychography, and (b) HAADF. The weak Z contrast of multislice ptychography results in similar contrast from the Sr and Ti-O columns, and weaker contrast from the O columns, reflecting the choice of the size of the atoms in (a). In (b), the oxygen atoms are completely removed, and the atoms in the Ti-O column have half the size as the Sr atoms to mimic the contrast of HAADF imaging. The two patterns visually appear different due to the different sizes assigned to the atoms, a distinction that is also evident in the experimental moiré patterns.

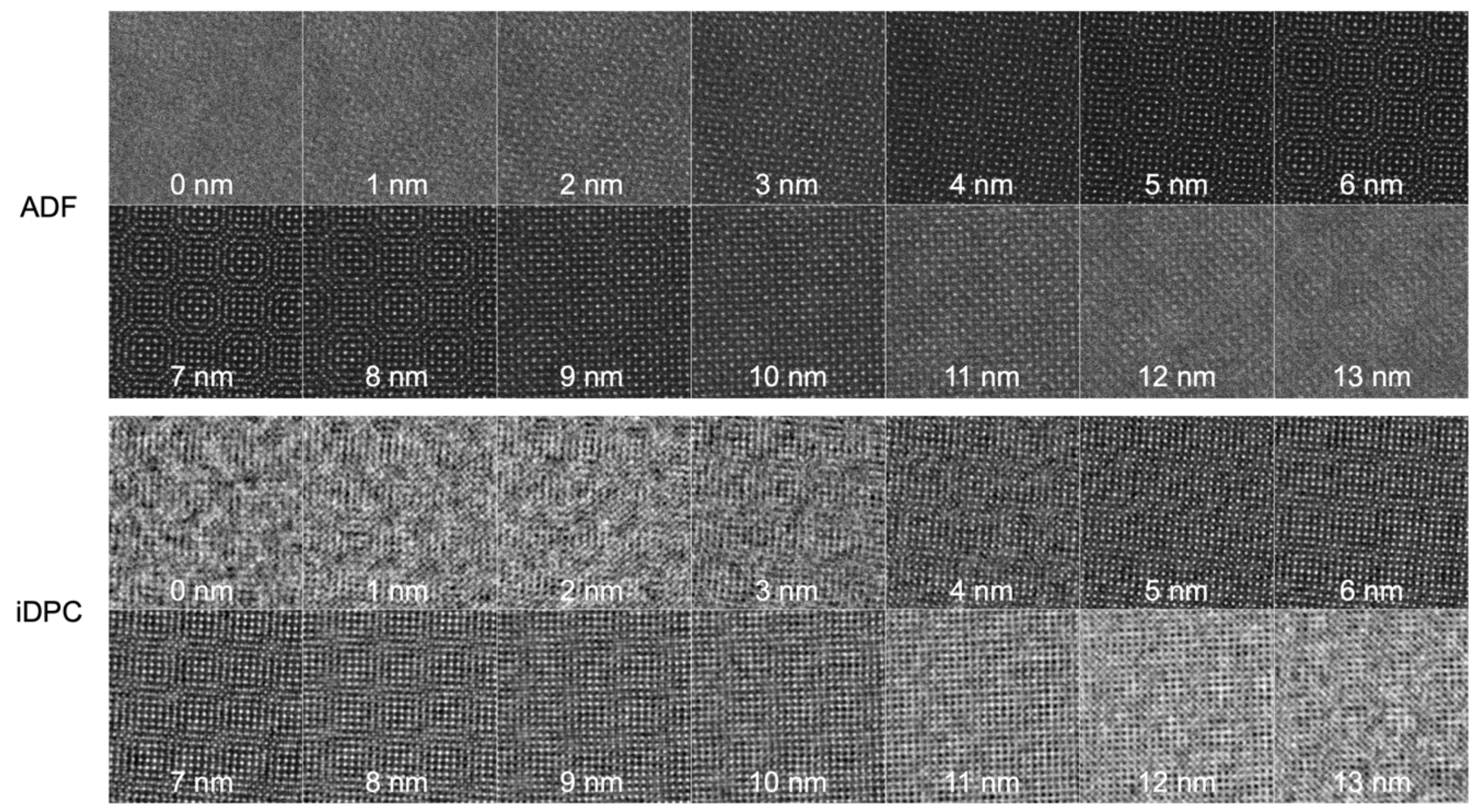

**Supplementary Fig. 9** | Montage of the images acquired with ADF and iDPC imaging in an optical through-focal series for Sample 1. The field of view shown in the above images is around 7.36 nm x 7.36 nm. Both ADF and iDPC images show good contrast only for a limited range of defocus.

**Supplementary Fig. 10** | Montage of different slices in the multislice ptychographic reconstruction of Sample 1 showing the sample structure at different depths in the beam direction.

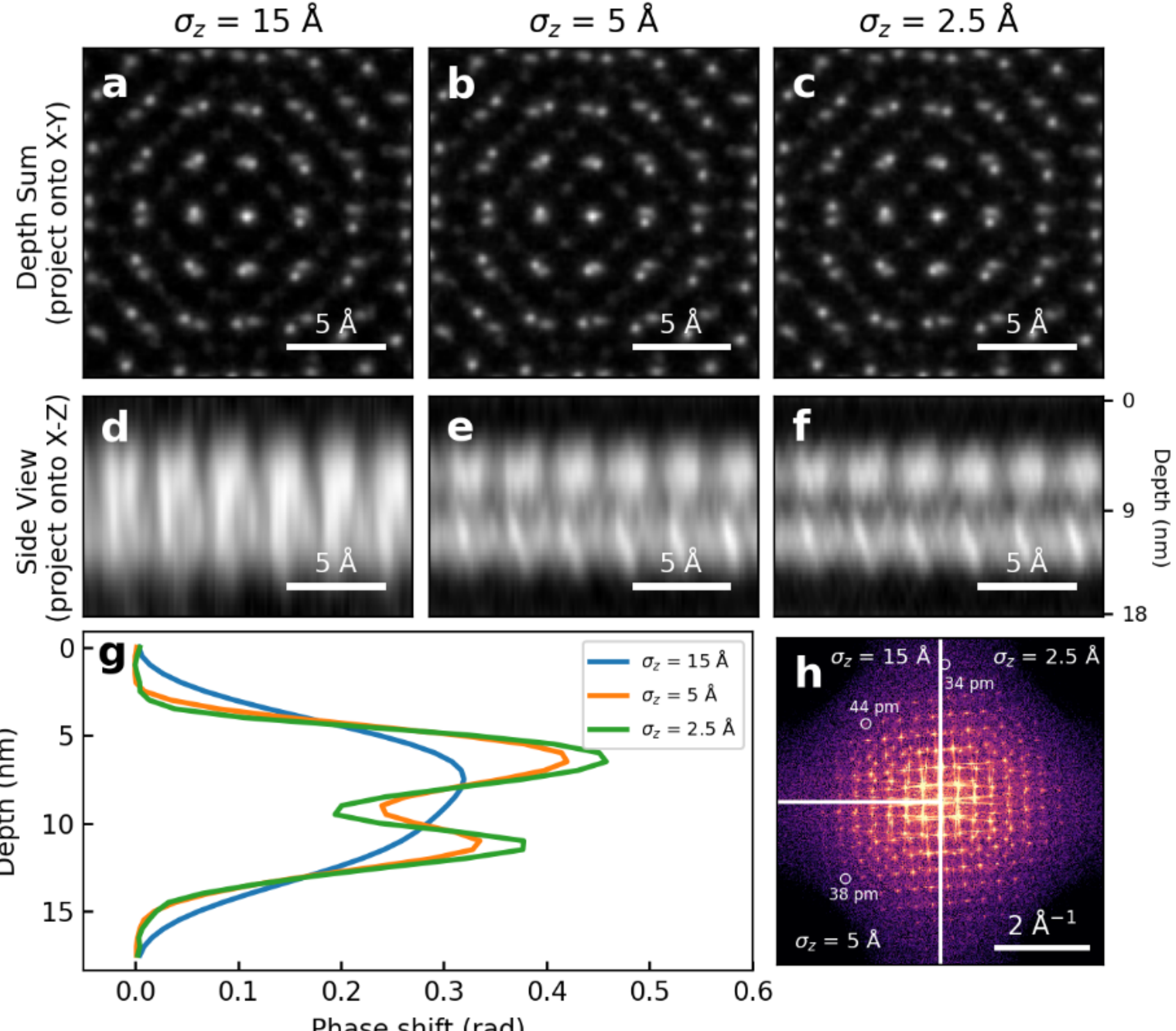

**Supplementary Fig. 11 |** Effect of depth regularization on multislice electron ptychographic reconstruction of twisted bilayer $SrTiO_3$ (STO) using PtyRAD. (a-c) Depth-summed projected phase images reconstructed with different depth-regularization strengths, parameterized by the Gaussian standard deviation $\sigma_z$ along the beam (z) direction: 15 Å, 5 Å, and 2.5 Å. (d-f) Corresponding side views of the reconstructed 3D phase volumes, shown as X-Z projections, illustrating clear inter-layer gaps with decreasing $\sigma_z$. (g) Line profiles along depth of an representative overlapping atomic column, highlighting the enhanced separation of the two STO membranes for smaller $\sigma_z$. (h) FFT power spectra of (a-c), showing increased information transfer as the depth regularization is relaxed.

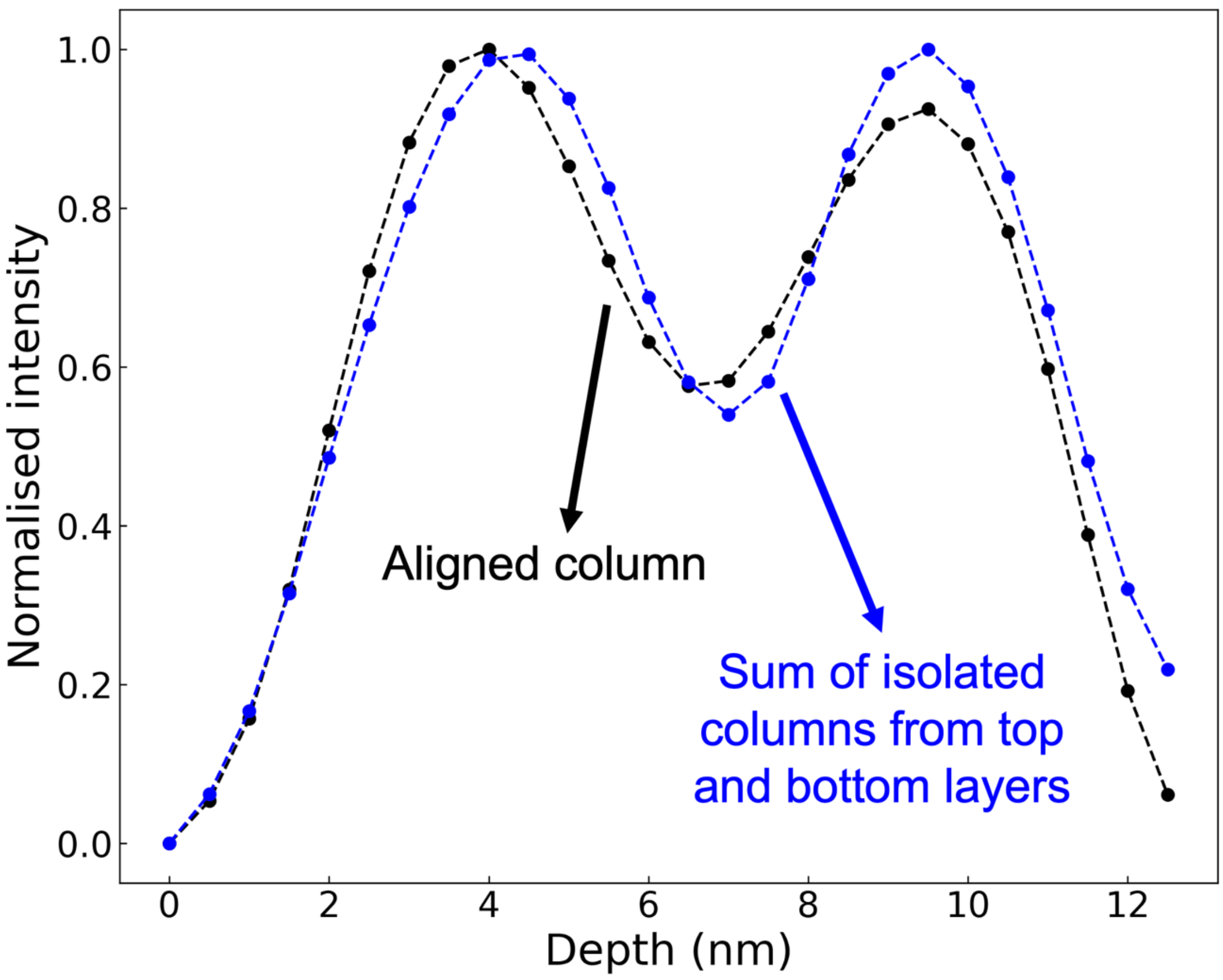

**Supplementary Fig. 12** | The depth profile shown in black is obtained by summing up two depth profiles corresponding to two separate atomic columns that are confined in the top membrane and the bottom membrane respectively. The depth profile shown in blue is measured along an atomic column that is aligned in both membranes. The close match between the two resulting depth profiles indicates the linearity of MEP as an imaging mode.

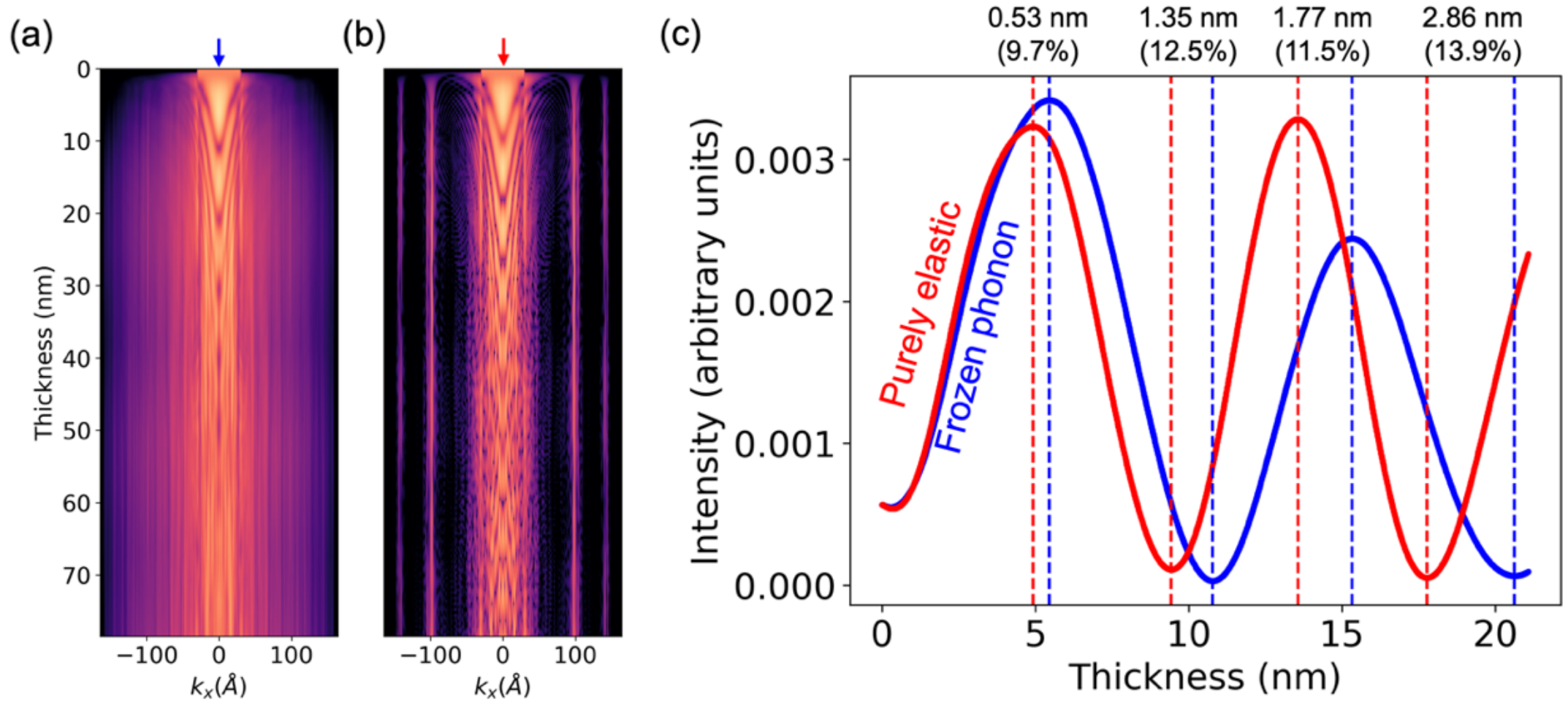

**Supplementary Fig. 13** | Comparison of channelling effects in diffraction patterns simulated with and without accounting for the effect of phonons with the probe focused on top of a Sr column in $SrTiO_3$. (a, b) The intensity profile through the diffraction pattern in the [100] direction is plotted as a function of thickness (a) accounting for electron-phonon scattering through the frozen phonon approximation[6] and (b) with a purely elastic model of diffraction. (c) Intensity variations of the center of the diffraction pattern marked with blue and red arrows in (a, b) plotted as a function of thickness. The addition of phonon scattering delays the onset of channelling induced minima and maxima in the intensity profile with respect to a purely elastic scattering model. For sample thickness in the range of 10-15 nm relevant to this study, this offset in thickness between the purely elastic model and the frozen phonon model is around 12%, consistent with the observed underestimation of thickness in the MEP reconstruction of experimental data. The diffraction patterns are simulated using abTEM[5] package with a probe semi-convergence angle of 30 mrad and beam energy of 300 keV.

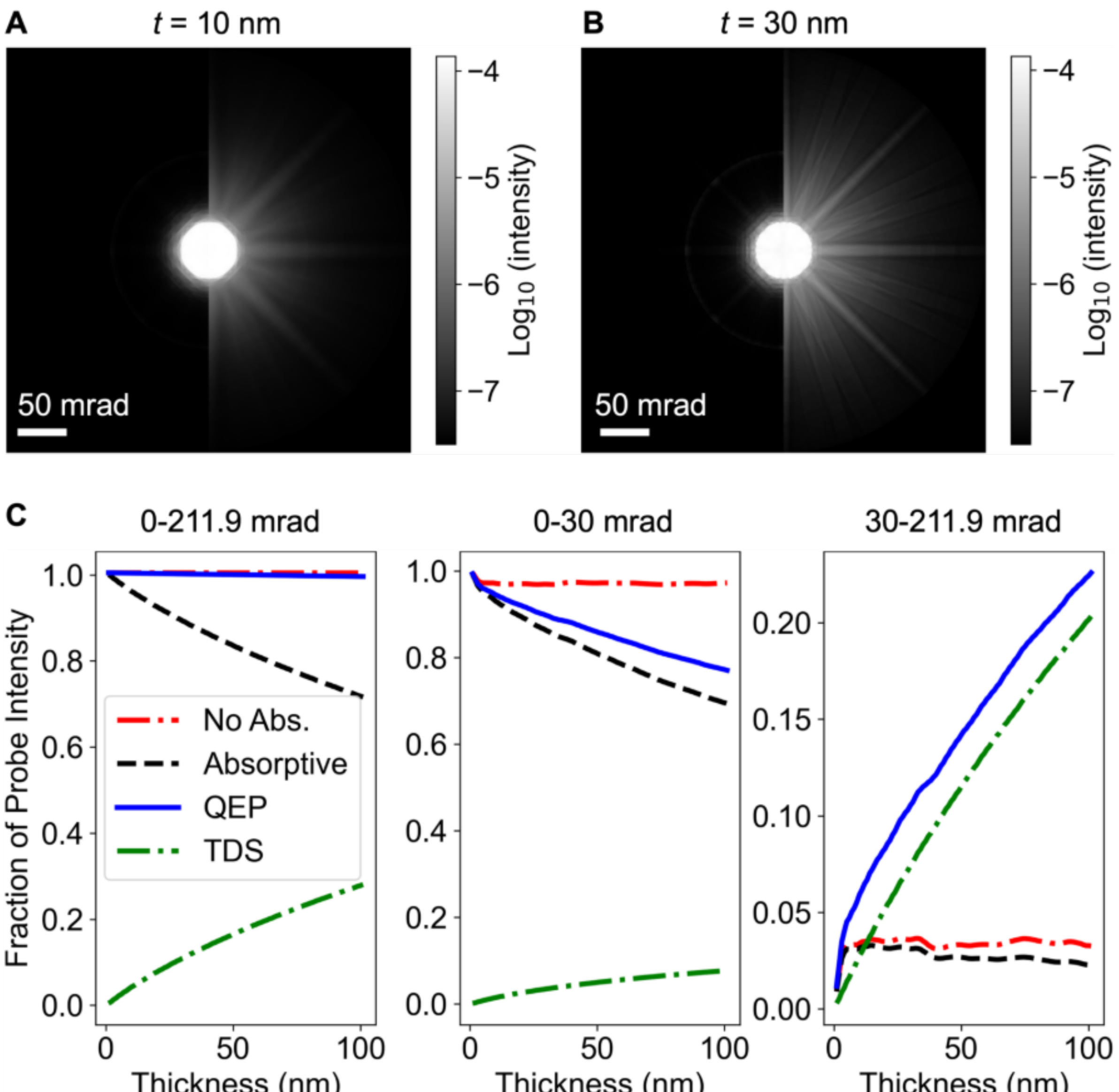

**Supplementary Fig. 14** | Intensity of position-averaged CBED patterns for $SrTiO_3$ along <001> as a function of thickness. (A, B) PACBED patterns were simulated using absorptive potential (left) and frozen phonon (right) for sample thicknesses t = 10 nm (A) and t = 30 nm (B), respectively. The simulations used a probe semi-convergence angle α = 30 mrad, and 1000 Monte Carlo steps for each probe position for the frozen phonon calculation. (C) PACBED total intensity as a function thickness for absorptive potential (Absorptive, black dashed curve), frozen phonon (QEP, blue curve), and thermal diffuse scattering (TDS, green dashed curve) in terms of fraction of the incident probe intensity. The intensities were integrated between scattering angles of 0-211.9 mrad (left), 0-30 mrad (middle, bright-field disk), and 30-211.9 mrad (right, dark-field).

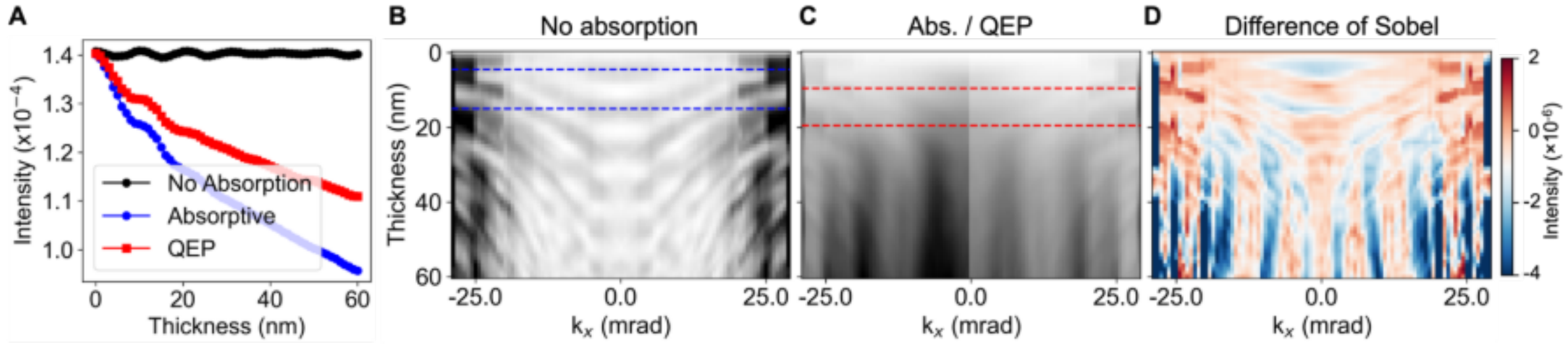

**Supplementary Fig. 15** | Pendellösung fringes of $SrTiO_3$ [001] PACBED patterns. (A) Bright-field intensity extracted at k = 0 as a function of specimen thickness for three scattering models: purely elastic scattering (no absorption), absorptive potential, and frozen phonon (QEP). Inclusion of thermal diffuse scattering leads to a monotonic reduction of the central-disk intensity and damping of oscillatory behavior with increasing thickness. (B,C) Thickness-dependent PACBED intensity maps along $k_y = 0$ for (B) no absorption and (C) absorptive potential (left) or frozen phonon (right). Dashed lines mark the first two intensity oscillations with increasing thickness: blue for the no absorption case (t = 4.5 and 15 nm) and red for the absorptive/QEP case (t = 9.5 and 19.5 nm). (C) Difference map obtained from Sobel-filtered images of (B) and (C), highlighting changes in extinction distances.

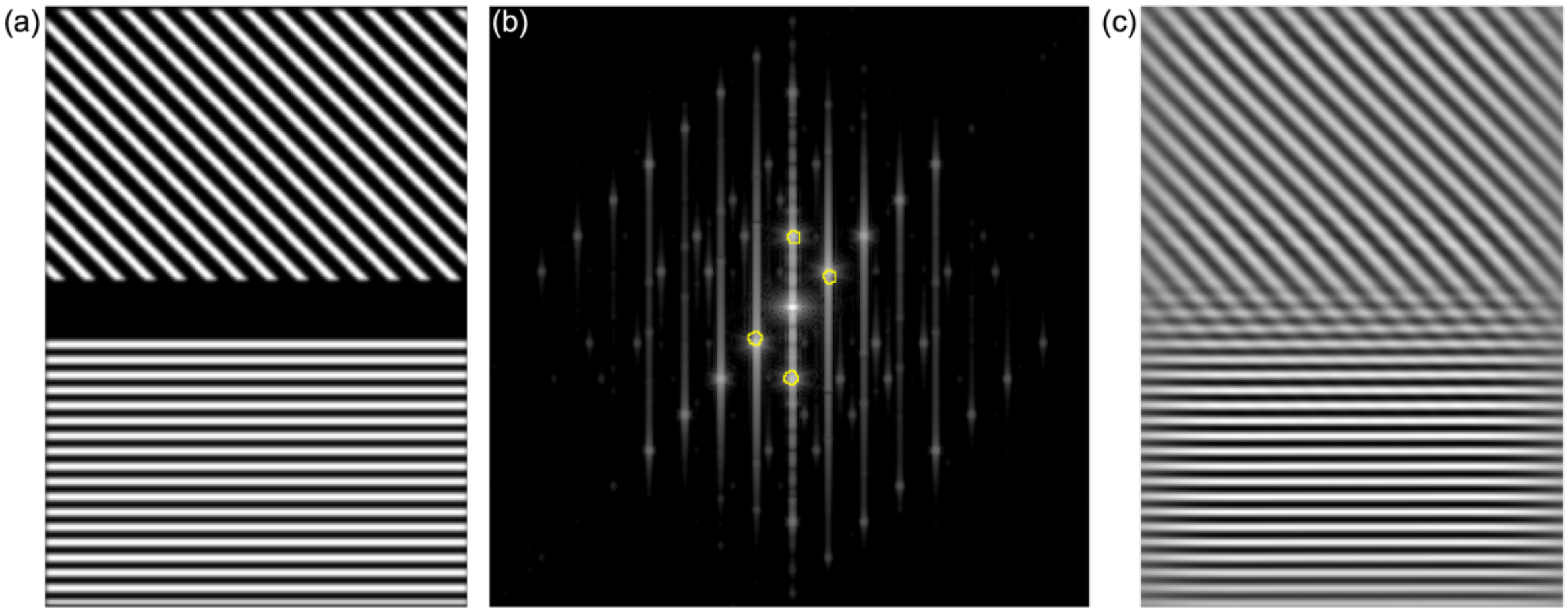

**Supplementary Fig. 16** | Illustration of how Fourier filtering can introduce artificial atomic planes. (a) An image with two sets of parallel lines with a gap in between made to mimic a gapped heterostructure. (b) Fourier transform of the image in (a) with yellow circles around the first order peaks that are used as masks to create the Fourier-filtered image in (c). The Fourier-filtered image shows artificial planes in the gap due to the real-space coarsening from the finite size of the mask in Fourier space. If filtering/denoising is needed for quantitative analysis, unbiased simple filters like isotropic Gaussian blurring are probably the safest.

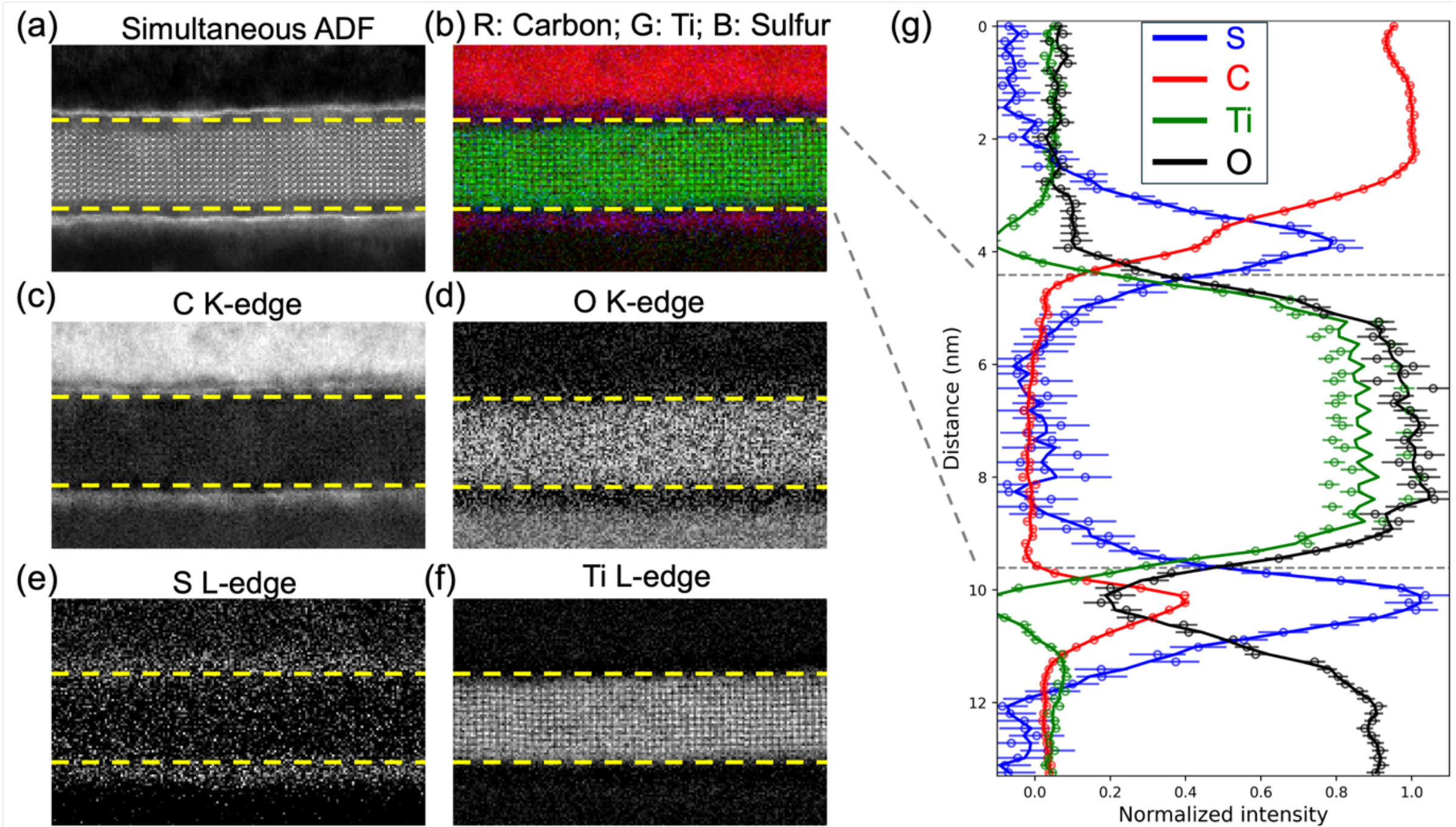

**Supplementary Fig. 17** | (a) ADF image of the $MoS_2$ (monolayer) / $SrTiO_3$ / $MoS_2$ (monolayer) heterostructure taken simultaneously with the EEL spectra. (b) Composite elemental map of carbon (red), titanium (green) and sulphur (blue) generated from (c, e, f). (c-f) Elemental maps of carbon, oxygen, sulphur, and titanium respectively calculated from the EEL spectra. (g) Averaged line profile across the interface calculated from the elemental maps in (c-f). The absence of a prominent carbon peak between the sulphur peaks from $MoS_2$ and the Ti peak from $SrTiO_3$ indicates that there is no interfacial contamination layer.

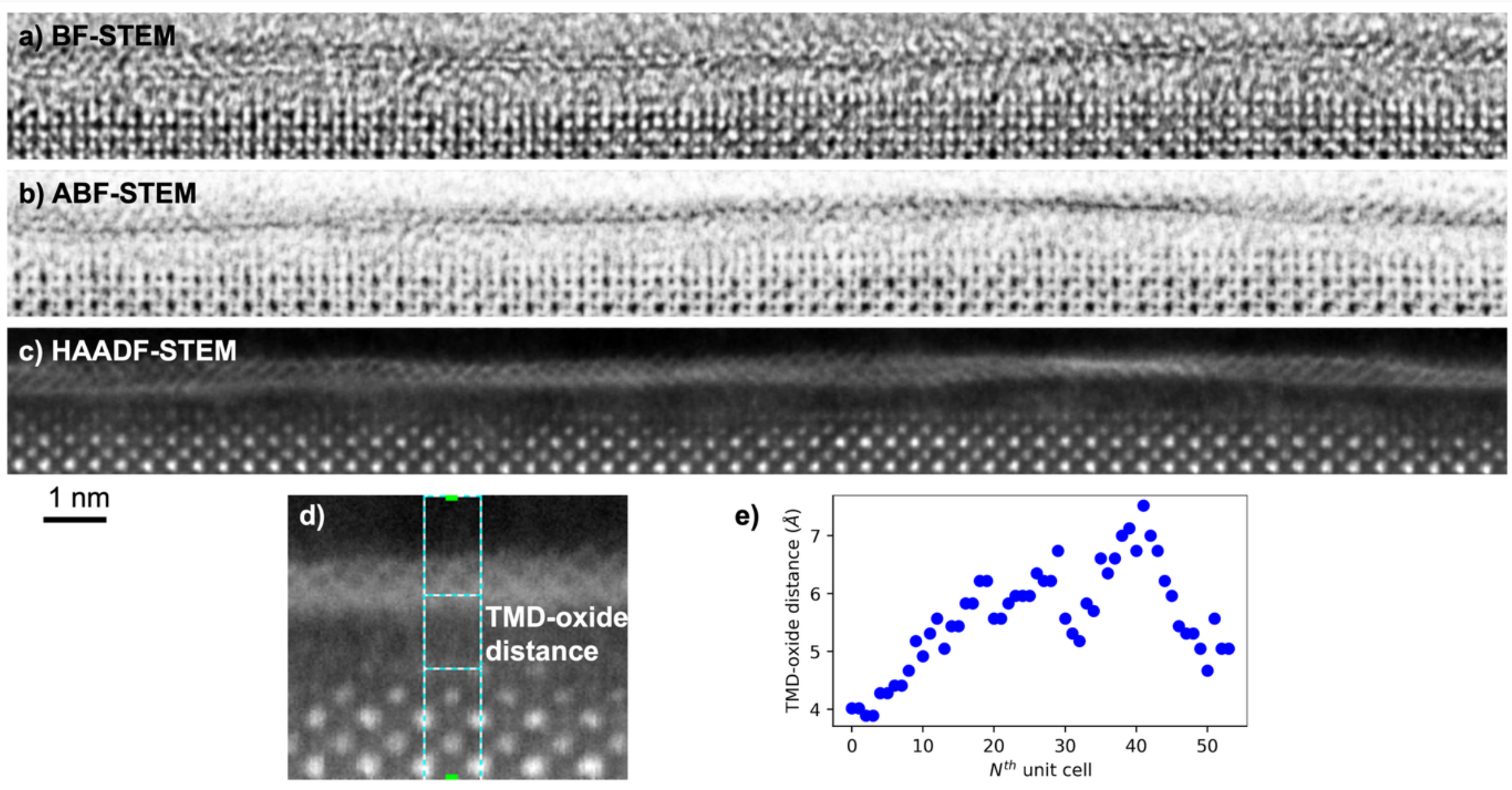

**Supplementary Fig. 18** | (a, b, c) show the BF, ABF, and HAADF images of the top TMD-oxide interface. d) Schematic showing the measurement of the projected TMD-oxide interfacial distance for every unit cell of the oxide. The measured values of the interfacial distance are plotted in (e).

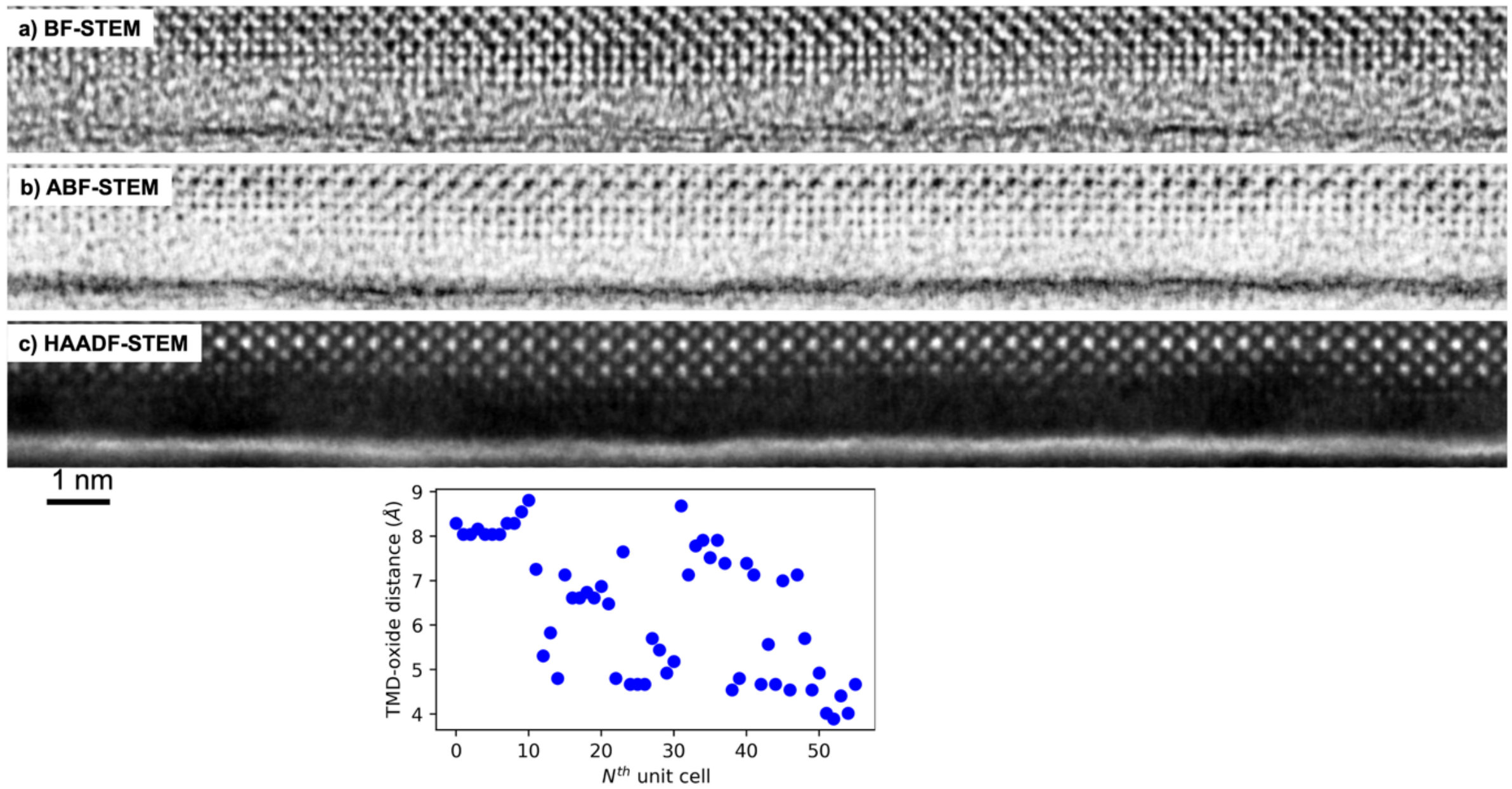

**Supplementary Fig. 19** | (a, b, c) show the BF, ABF, and HAADF images of the bottom TMD-oxide interface. d) Schematic showing the measurement of the projected TMD-oxide interfacial distance for every unit cell of the oxide. The measured values of the interfacial distance are plotted in (e).

**Supplementary Note 1**

To investigate the interplay between sample thickness, gap size, and the effect of depth-resolution-limited blurring, we derive (1) an analytical model with semi-infinite slabs, and present (2) numerical simulations with finite thickness as shown in Supplementary Fig. 20. We find that in typical experiment conditions of MEP, discernible contrast dip $C$ can be qualitatively interpreted as a gap, while the quantification of exact gap size remains challenging due to uncertainties in sample thickness, surface roughness, and spatial variability in depth resolution.

**(1) Analytical model with semi-infinite slabs**

The dip in intensity between two semi-infinite slabs, separated by a gap, $G$, and blurred by a normalized Gaussian with a standard deviation $\sigma$ can be modelled analytically as

$$F(x) = 1 + \frac{1}{2}\left[\operatorname{erf}\left(\frac{x-\frac{G}{2}}{\sqrt{2}\sigma}\right) - \operatorname{erf}\left(\frac{x+\frac{G}{2}}{\sqrt{2}\sigma}\right)\right] \quad (1)$$

Replacing $G$ with dimensionless parameter $\tilde{G} \equiv (G/\sigma)$ gives

$$F(x) = 1 + \frac{1}{2}\left[\operatorname{erf}\left(\frac{x}{\sqrt{2}\sigma} - \frac{\tilde{G}}{2\sqrt{2}}\right) - \operatorname{erf}\left(\frac{x}{\sqrt{2}\sigma} - \frac{\tilde{G}}{2\sqrt{2}}\right)\right] \quad (2)$$

The minimum occurs at $x = 0$,

$$F_{\min} = F(0) = 1 - \operatorname{erf}\left(\frac{\tilde{G}}{2\sqrt{2}}\right) \quad (3)$$

and the contrast dip, $C$, is simply $C = \operatorname{erf}\left(\frac{\tilde{G}}{2\sqrt{2}}\right)$. In the limit of very small $\tilde{G}$, this can be approximated as $C \approx \frac{\tilde{G}}{\sqrt{2\pi}}$ using the relation of $\operatorname{erf}(x) \sim \frac{2}{\sqrt{\pi}} x$, valid for $C \lesssim 0.2$ with approximation error around $0.01C$. Writing this in terms of the smallest gap, $G_{min}$, that is detectable for a given contrast dip, we get $G_{min} = \sqrt{2\pi}\ \sigma C$. For a 20% contrast dip and a typical $\sigma = 1$ nm for MEP (a 2.36 nm FWHM Gaussian depth blur), we would get $d_{min} \approx 0.50$ nm. However, this model breaks down for thin samples as the intensity would never reach the bulk level, making quantification more challenging. Therefore, we present numerical simulations in Supplementary Fig. 20 (b-c).

**(2) Numerical simulation with finite slabs**

The numerical simulations are done with similar setup, except with an additional variable being the slab thickness, $L$, and both slabs have the same thickness. Following the analytical model, we define dimensionless unit $\tilde{\sigma} \equiv (\sigma/G)$ and $\tilde{L} \equiv (L/G)$, using gap size $G$ as the unit length. Panel b clearly shows that with finite slab thickness ($\tilde{L} = 5$, or equivalently two 5 nm slabs separated by a 1 nm gap), the peak intensity is becoming lower than the bulk value with increasing $\tilde{\sigma}$. When $\tilde{\sigma} = 2$ (or equivalently a FWHM of 4.71 nm blurring a 1 nm gap, close to the depth resolution of through-focus ADF-STEM), the dip becomes nearly invisible. Note that the perceived dip is also a function of sample thickness. Panel c shows that a minimal $\tilde{L} = 5$ is needed to preserve the bulk intensity when $\tilde{\sigma} = 1$. For larger $\tilde{\sigma}$, the required $\tilde{L}$ is also larger. In summary, a discernible gap is evidently a sign of gap existence, however, quantification of gap size relies on accurate estimation of challenging parameters including sample thickness and depth resolution, which contain significant spatial variability, and can be further complicated by surface roughness.

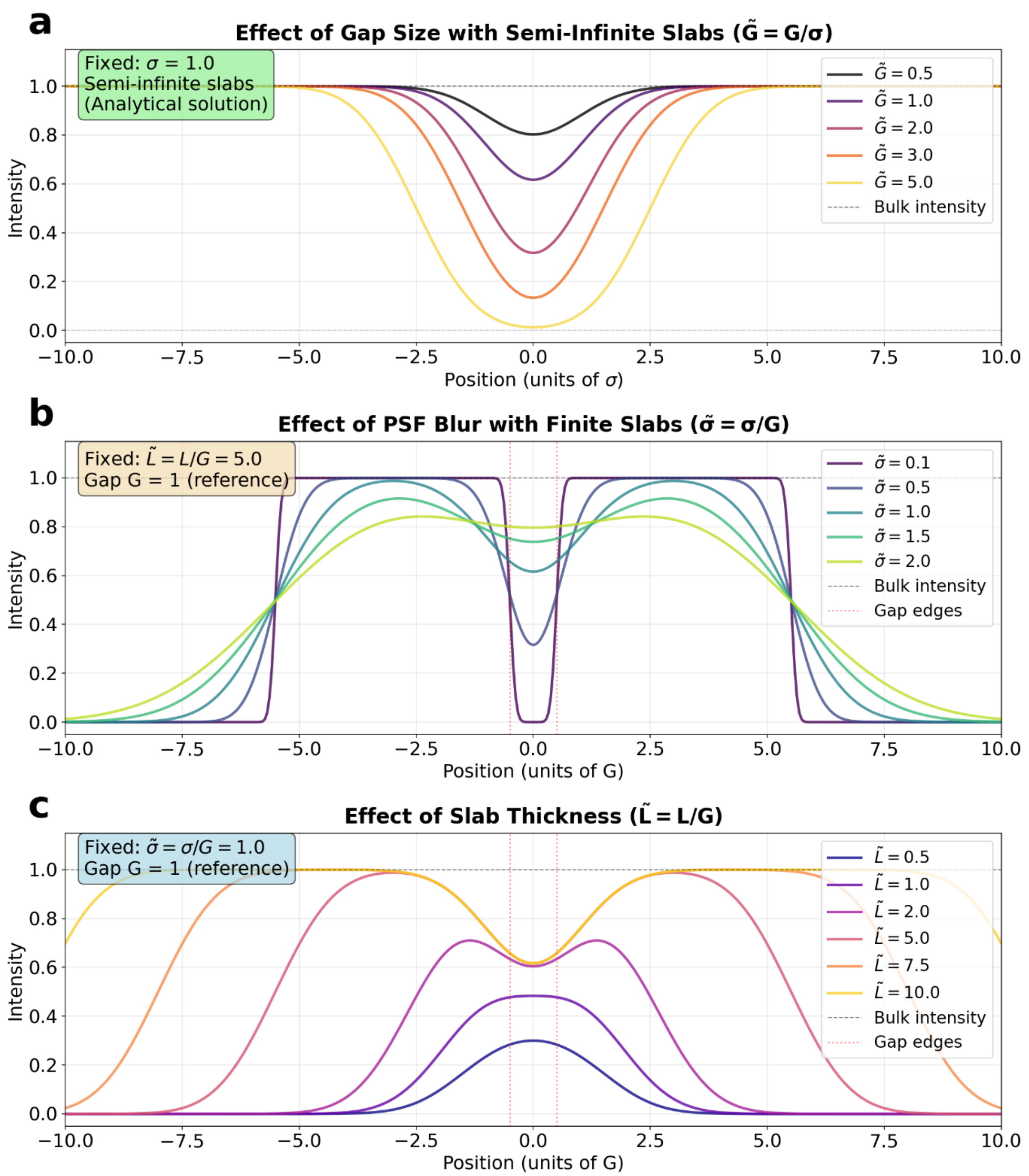

**Supplementary Fig. 20 | Effect of gap size, depth resolution, and slab thickness to visibility of the gap.** (a) Analytical model of two semi-infinite slabs separated by G, convolved with a Gaussian blur (standard deviation = $\sigma$), emulating the line profile with instrumental blurring. The contrast dip $C$ can be approximated as $C \approx \frac{\tilde{G}}{\sqrt{2\pi}}$, where $\tilde{G} \equiv (G/\sigma)$. (b) Numerical simulation of finite-thickness slabs with fixed lengths $\tilde{L} \equiv (L/G) = 5$ and varying depth resolution blur $\tilde{\sigma} \equiv (\sigma/G)$. Note that the peak intensity drops significantly when $\tilde{\sigma}$ is large. (c) Numerical simulation of finite-thickness slabs with fixed $\tilde{\sigma} = 1$ and varying slab thickness $\tilde{L}$, showing a minimal $\tilde{L}$ is required to reach bulk intensity.

**Supplementary Note 2**

We also present a mechanics estimate comparing adhesion energies and bending energies based on previous work[7,8]. The critical radius of curvature ($R_{crit}$) at which the bending energy and adhesion energy are in equilibrium is given by:

$$R_{crit} = \sqrt{B/\Gamma} \quad (3)$$

Here, B is the bending stiffness and Γ is the areal energy density of adhesion. The adhesion energy per unit area for $MoS_2$ on $SiO_2$ is estimated to be 200 ± 50 mJ $m^{-2}$ (1.25 eV $nm^{-2}$) which is likely similar to that for $MoS_2$ on oxygen-terminated $SrTiO_3$. The bending stiffness for $MoS_2$ is B = 10.15 ± 1.4 eV per monolayer. Using these values, the estimated critical bending radius for monolayer $MoS_2$ on $SrTiO_3$ is 2.8 ± 0.7 nm. This means that the MoS2 film cannot conform to roughness details with radii of curvature below 3-4 nm, which is consistent with the length scale of bends we see in our Fig. 6.

For finite-thickness oxide membranes, from Timoshenko beam theory[9], the bending energy per unit area, u, in terms of the membrane thickness t, radius of curvature R and bulk modulus E is $u = (Et^3)/(24R^2) = B/R^2$. For $SrTiO_3$ with a thin-film "bulk" modulus of ~200 GPa[10] the bending stiffness is $B \approx 50t^3$ with B in eV, and t in nm. The van der Waals energies for oxygen-terminated $SrTiO_3$ will be similar to the $MoS_2/SiO_2$ case (O vs S), so for a 5-10 nm thick $SrTiO_3$ membrane on $SrTiO_3$, so the critical radius will be 25-70 x larger than for $MoS_2$ on $SrTiO_3$, i.e. we would expect $SrTiO_3$ membranes to have trouble conforming to features smaller than 70-200 nm in size for van der Waals forces, and 7-20 nm for direct chemical bonds where adhesion energies are roughly 100x larger.